\documentclass[notitlepage,prd,aps,longbibliography,twocolumn]{revtex4-1}
\usepackage{amsmath,amsfonts,amsthm,bm}\usepackage{bbm}\usepackage{Abbreviations}
\usepackage{dsfont}
\usepackage{color}
\begin{document}
\title{
Casimir and Casimir-Polder forces with dissipation from first principles}
\author{M. Bordag\footnote{bordag@uni-leipzig.de},
\footnotesize{{\sl Institut f\"ur Theoretische Physik, Universit\"at Leipzig, Germany.}}}
\date{December 10, 2017
}

\begin{abstract}
We consider Casimir-Polder and Casimir forces with finite dissipation by coupling heat baths to the dipoles introducing, this way, dissipation from 'first principles'. We derive a representation of the free energy as an integral over real frequencies, which can be viewed as an generalization of the 'remarkable formula' introduced by Ford et. al. 1985. For instance, we obtain a nonperturbative representation for the atom-atom and atom-wall interactions. We investigate several limiting cases. From the limit $T\to0$ we show that the third law of thermodynamics cannot be violated within the given approach, where the dissipation parameter cannot depend on temperature 'by construction'. We conclude, that the given approach is insufficient to resolve the thermodynamic puzzle connected with the Drude model when inserted into the Lifshitz formula. Further we consider the transition to the Matsubara representation and discuss modifications of the contribution from the zeroth Matsubara frequency.
\end{abstract}
\maketitle   
\section{\label{T0}Introduction}
Casimir and Casimir-Polder forces are the basic interaction between single atoms and macroscopic bodies at separations beyond the direct overlap of the electronic wave functions. Their range stretches from a few nanometers untill macroscopic separations.  In  many areas of physics, and beyond, these are ubiquitous. Frequently, these are also called {\it dispersion forces} \cite{mahanty76}. These forces are due to the quantum electromagnetic interaction between polarizable dipoles formed by the atoms and  the polarization inside the bodies. There are several instructive explanations for these forces. One considers the influence of the dipoles or of the bodies  on the vacuum of the electromagnetic field, generalizing the notion of zero point energy known long ago in quantum mechanics. The other considers the fluctuations, quantum and thermal, of the electromagnetic field in the presence of the polarizable bodies. The first was pioneered by Casimir for ideally reflecting bodies \cite{casi48-51-793}, the second by Lifshitz \cite{lifs56-2-73} for polarizable bodies. In a modern view, both have the same origin -- the expectation value of the Hamiltonian (or the energy-momentum tensor) of the considered system taken  in the vacuum or in a thermal state. Thereby the latter is usually described by the Matsubara formalism.

The dispersion force between macroscopic bodies is most frequently described by the Lifshitz formula (LF) \cite{bara75-116-5}. All other cases can be obtained from here. For instance, for ideal conductors the original Casimir force follows, for rarified media the Casimir-Polder forces for atom-wall or atom-atom interactions follow (see \cite{BKMM} and citations therein). The LF is highly versatile for describing the interaction of real bodies as it uses the reflection coefficients of the bodies and the polarizability of the atoms as input. By using known theoretical models or experimental data for them, nearly any real situation can be handled. An impressive level of precision comparison between theoretical predictions and measurements has been reached \cite{klim09-81-1827}. In line with such progress, unresolved problems remained. On the theoretical side these manifest themselves, for instance, in a violation of the third law of thermodynamics (Nernst's heat theorem). As shown first in \cite{beze02-65-052113}, inserting the Drude model permittivity into the LF, for a dissipation parameter $\ga$, decreasing with temperature $T\to0$ faster than the first power, $\ga\sim T^\al\ \ (\al>1)$, at $T=0$ a non-zero entropy remains (see \cite{klim09-81-1827}, Sec. II.D, for details). On the experimental side, the use of the Drude model permittivity results in theoretical predictions, which are for some configurations in growing disagreement with measurements, see for instance \cite{bimo16-93-184434}. There are similar problems for insulators with dc conductivity. There were also several attempts to resolve these problems, but all were not convincing, eventually.

One of the most controversial points is the question about the validity of the LF with Drude permittivity inserted. A main part of this discussion concerns the assumption of thermodynamic equilibrium assumed in deriving the LF, which may be violated in case of dissipation of free charge carriers (electrons). On the other side, the LF is formulated in terms of reflection coefficients, which may be derived, or measured, independently. This way, it is not clear whether one is allowed to insert the Drude model permittivity, especially with a temperature dependent dissipation parameter, into the LF.

In past years, several attempts have been undertaken to reach a derivation of the LF with Drude permittivity from first principles. In fact, this is possible by attaching heat bathes (reservoirs) to the oscillators interacting with the electromagnetic field and allows to 'generate' a Drude permittivity in an effective equation for the electromagnetic field. As a result, the validity of the LF with Drude permittivity inserted, could be verified. An early work in this direction was \cite{bara75-116-5}; more recent are \cite{kupi92-46-2286}, \cite{rosa11-84-053813}, \cite{lomb11-84-052517}, \cite{lope17-28-025009}, \cite{brau17-190-237}.

In the present paper we consider resp. reconsider the heat bath approach. Thereby we give a direct derivation in simple terms of the involved fields and their Green's functions. We arrive at a simple and universal representation for the free energy of the considered configurations.
These are atom-atom, atom-wall and wall-wall configurations at some separation $a$. The first two are commonly named after Casimir and Polder \cite{casi48-73-360} and the latter corresponds to the Casimir effect \cite{casi48-51-793} and to the LF \cite{lifs56-2-73}. For all configurations we end up with a representation
\be F=\int_0^\infty \frac{d\om}{\pi}
    \left(\frac{\hbar \om}{2}+k_{\rm B} T \ln\left(1-e^{-\beta\hbar\om}\right)\right)
       \frac{\pa}{\pa \om}\delta(\om)
\label{1}\ee
for the free energy, where
\be \delta(\om)=\frac{-1}{2i}{\rm tr}\ln\frac{L(\om)}{L(\om)^*}
\label{2}\ee
(for details see below) is a kind of phase, the specific expression for   depends  on the configuration considered. In the case of no dissipation, $\delta(\om)$ turns into the scattering phase shift (in the case of a continuous spectrum).
{For a recent example see eq. (96) in \cite{D17-1}.}

In fact, the representation \Ref{1} for the free energy can be viewed as a kind of generalization of the 'remarkable formula', derived in \cite{ford85-55-2273} for a single oscillator, where the role of $\delta(\om)$ is played by a generalized susceptibility resp. the imaginary part of a Green's function. In the literature there are more generalizations and applications; one of the first can be found in \cite{bara75-116-5}, eq. (16), and a more recent one in \cite{intr12-86-062517}, which is in terms of modes.

As concerns the structure of eq. \Ref{1}, it must be mentioned that in case of no dissipation the vacuum energy (at $T=0$) can be expresses as a sum over modes,
\be E_0=\frac{\hbar}{2}\sum_{J} \om_J,
\label{3}\ee
where the $\om_J$ are the eigenfrequencies of the $J$-th mode (dropping any ultraviolet  regularization). In case of a continuous spectrum, the sum must be substituted by a corresponding integral. In equilibrium, and with no losses, these frequencies are real. However, in case of dissipation, these have an imaginary part and so does the vacuum energy \Ref{3}, indicating an instability. In this context it is interesting to remark that \Ref{1} is an integral over a real variable, $\om$, having the meaning of a frequency.  However, this frequency is not related to the frequencies $\om_J$ in \Ref{3}. Also, the physics behind \Ref{1} is different. Eq. \Ref{1}, is derived from the heat bath approach. It assumes the coupling of each oscillator (mode) to a reservoir. As a result, the motion of these oscillators is damped, but the damping is in equilibrium with the driving Langevin force. There are, in addition, the eigenmodes $\om_J$, but these die out with time and will not contribute in equilibrium. For this reason it would be misleading to call \Ref{1} a 'sum over modes'.

Further in the present paper we check the limit of vanishing temperature, $T\to0$, and find in all configuration a decrease of the temperature dependent part of the free energy $\Delta_TF\sim T^2$ at least, which ensures non-violation of third law of thermodynamics. This holds also for the case if the intrinsic frequency of the oscillators is zero. It must be mentioned that the above mentioned violation of the third law appears if the dissipation parameter itself depends on temperature, $\ga(T)$, and decreases sufficiently fast. We show, that this violation happens in the heat bath approach too. In other words, the heat bath approach does not cure the violation problem with thermodynamics either. The point is that the dissipation parameter, within the heat bath approach, does not depend on temperature by construction (it may only depend on frequency) and any its temperature dependence is alien to this approach. It remains a kind of phenomenological input and does not follow from the first principles approach taken here. Thus, the solution of the mentioned problem is beyond the approach taken in this paper and further work is necessary.

For the Casimir-Polder configurations we obtain formulas which are non-perturbative in the polarizability $\al$ of the dipole(s). The commonly considered case of large separations or small polarizability is then obtained by expansion in powers of $\al$ or $1/a$. This expansion has a finite radius of convergence which is determined by the onset of instability when decreasing the separation $a$. This instability corresponds to spontaneous creation of photons. We assume that the corresponding states are occupied and exclude them from the statistical ensemble when doing the thermal averaging. A non-perturbative approach to the Casimir-Polder force, using a diagonalisation of the corresponding Hamiltonian, was developed in \cite{pass12-85-062109}, where also possible applications are discussed. Another nonperturbative approach to the Casimir-Polder force was taken in \cite{berm14-89-022127}, where also a discussion of the instability was included.
In electrodynamics, for the interaction of atoms, the instability would appear for separations $a$ of order of the size of atoms, $a\sim\al^{-1/3}$, which is beyond the applicability of the dipole approximation. To conclude this topic, we mention that such kind of instability does not appear in the LF.

Further, in the present paper, we consider the transition from the representation \Ref{1} of the free energy to the Matsubara representation. This can be done by a kind of  Wick rotation. For parameters, keeping the considered configuration below criticality  and giving $L(0)$ a finite value, we obtain in each case the usual Matsubara representation. For the other cases we observe a transmutation of the contribution from the zeroth Matsubara frequency into a logarithmic term and an additional term in case of criticality. Thereby we pay special attention to the case of vanishing intrinsic frequency of the oscillators and discuss the resulting modifications of the results.

Finally in the present paper, we consider the relation of the considered model with the plasma model. As known, the permittivity of the plasma model follows from that of the Drude model by formally putting the dissipation parameter to zero, $\ga=0$, whereas the free energy does not turn into that of the plasma model for $\ga\to0$. The difference can be traced back to contributions in \Ref{1} with $\om$, which are smaller than the frequencies in the spectrum of the modes of the plasma model.

Throughout the paper three and two dimensional vectors and matrixes are denoted by bold letters. For instance, the coordinate is $\x=(x,y,z)=(\x_{||},z)$, where $\x_{||}$ is a vector in the plane  of an interface.
We use units with $\hbar=c=k_{\rm B}=1$.
\renewcommand{\hbar}{} 
\\
We reserve the notation $\delta(\om)$ for what is called 'phase' in this paper. Spatial delta functions are always denoted by $\delta^{(1)}(x)$ or
$\delta^{(3)}(\x)$, in dependence on the dimensionality.
%
\section{\label{T2}The model, basic formulas and configurations}
In this section we collect the basic formulas for the considered model and configurations.
\subsection{\label{T2.1}The model}
The model consists of  polarizable atoms and their interaction with the electromagnetic field and with heat baths. From the Maxwell equations we have
\be \left(\pa_t^2-\Delta+\bnabla\circ\bnabla\right)\E(t,\x) = 4\pi\pa_t\, \j(t,\x)
\label{2.1}\ee
as equation for the electric field in dyadic notation with  the current density $\j(\x)$ as source. The magnetic field follows with ${\rm rot}\E=-\pa_t \B$.
The atoms are described by point dipoles and the interaction is taken in dipole approximation.
The $i$-th atom is described by a dipole with the charge $e$ at location $\x=\a_i$ and the displacement $\bxi_i(t)$ of its charge. Its dipole moment is $\bm{p}_i(t)=e\bxi_i(t)$ and this dipole is  equipped with the dynamics of a harmonic oscillator,
\be  m\left(\pa_t^2+\ga\pa_t+\Om^2\right)\bxi_i(t) = e\E(t,\a_i)+\F_i(t),
\label{2.2}\ee
where $\Om$ is the intrinsic frequency of the oscillator, $\ga$ is the damping constant and $\F(t)$ is the Langevin force. The current density generated by the atoms is
\be \j(t,\x)=e\pa_t\sum_i\bxi_i(t)\delta^{(3)}(\x-\a_i).
\label{2.3}\ee
The considered system has a   Lagrangian
\bea {\cal L} &=& \int d\x\,\frac{1}{8\pi}\left(\E(t,\x)^2-\B(t,\x)^2\right)
\label{2.4}\\&&        +\sum_i\frac{m}{2}\left(\bm{\dot{\xi}}_i(t)^2-\Om^2\bxi_i(t)^2\right)
        +\sum_i \bm{p}_i(t)\E(t,\a_i),
\nn\eea
where we did not show the heat bath part, and the classical energy is given by
\be E=E_{\rm ED}+E_{\rm dipole},
\label{2.5}\ee
with
\bea  E_{\rm ED} &=& \frac{1}{8\pi}\int d^3\x\left(\E(t,\x)^2+\B(t,\x)^2\right),
    \nn\\
    E_{\rm dipole} &=& \sum_i\frac{m}{2}\left(\dot{\bxi}_i(t)^2+\Om^2\bxi_i(t)^2\right),
\label{2.6}\eea
(see \cite{Akhiezer1985}, p.334, or \cite{milton98}, Eq.(9.37)). The heat bath part, which we do not show here, and how it results in the damping term and the Langevin forces, was discussed in detail in the scalar example in \cite{D17-1} and also elsewhere in literature, see for example \cite{rosa10-81-033812}.
From the above Lagrangian (including the heat bath part), the equations of motion, Eqs. \Ref{2.1} and \Ref{2.2}  for field and the oscillators, and the equations for the bath variables as well,  can be derived. This way, the approach starts from 'first principles'.

Since our setup keeps translation invariance in time, it is meaningful to apply Fourier transform in time. For the electric field we have
\bea \E(t,\x)&=&\int_{-\infty}^\infty\frac{d\om}    {2\pi}e^{-i\om t}\tE_{\om}(\x),
    \nn\\ \tE_{\om}(\x)&=&\int_{-\infty}^\infty dt\,    e^{i\om t}\E(t,\x),
\label{2.10}\eea
and similar for all other time dependent quantities. The transformed quantities are always denoted by a tilde. After Fourier transform \Ref{2.10}, we get from Eq. \Ref{2.1} for the electric field and from Eq. \Ref{2.2} for the displacements,  the set of equations,
\begin{widetext}
\bea    (-\om^2-\Delta+\bnabla\circ\bnabla)\tE_\om(\x) &=&
4\pi e\om^2\sum_{i}\txi_{\om, i}\delta^{(3)}(\x-\a_i),
\nn\\
m(-\om^2-i\ga\om+\Om^2)\txi_{\om, i}  &=& e\tE_\om(\a_i)+\tF_{\om,i},
\label{2.11}\eea
\end{widetext}
which is the starting point for the following.

As said above, we assume an individual continuous heat bath coupled to each oscillator. As known, a heat bath, or reservoir,  can be represented as a continuous set of harmonic oscillators. In simple terms this procedure is described in detail for a one dimensional case in \cite{D17-1}. In \cite{ford88-37-4419}, the most general formulation and its relation to the oscillator model is discussed along with different couplings between bath and oscillator. The procedure goes as follows. First one solvers the Heisenberg equation of motion of the bath operators. As a result, in the equations of motions of the dipole oscillators, Eqs. \Ref{2.11}, second line, the friction term, $-i\ga\om$, and the Langevin force, $\tF_{\om,i}$, in the right side, appear. The damping parameter, $\ga$, may depend on frequency, being a positive function, but we restrict ourselves to a constant $\ga>0$. The Langevin force, $\tF_\om$, can be expressed in terms if the heat bath operators. Here we need only their thermal averages,
\be    \langle \tF_{\om,i} \tF_{\om',j} \rangle  =
 \frac{\ga m \om}{\pi}\delta(\om+\om') \delta_{ij} \coth\frac{\beta\om}{2}.
\label{2.12}\ee
Below we consider the following thermodynamic potentials. The internal energy $U$ is the thermal average of the Hamiltonian \Ref{2.3},
\be U=\langle H\rangle.
\label{2.13}\ee
It will be calculated by solving Eqs. \Ref{2.1} and \Ref{2.2} for the corresponding configuration, expressing the electric field and the displacement fields in terms of the Langevin forces and using the averages \Ref{2.12}. Further, by means of the thermodynamic relations,
\be  U=\frac{\pa}{\pa \beta}(\beta F),\qquad S=-\frac{\pa}{\pa T}F,
\label{2.14}\ee
the free energy $F$ and the entropy $S$ can be calculated.
\subsection{\label{T2.2}Solutions for the electric field and the displacements}
In this subsection,  we solve the system \Ref{2.11}. As discussed in \cite{D17-1}, two ways are possible. The first starts with solving first the equation for the displacements, second line in \Ref{2.11}, which then will be inserted into the equation for the electric field, first line in \Ref{2.11}. The second way, which we will follow now, goes in inverse order. We start by solving the equation for the electric field from \Ref{2.11},
\be \tE_\om(\x)=4\pi e\om^2 \sum_i\G^{(0)}_\om(\x-a_i) \txi_{\om,i},
\label{2.15}\ee
where we introduced the free-space Green's function $\G^{(0)}_\om(\x)$, obeying the equation
\be \left(-\om^2-\Delta+\bnabla\circ\bnabla  \right)\G_\om^{(0)}(\x-\x')=\delta^{(3)}(\x-\x').
\label{2.16}\ee
It has the representation
\be \G_\om^{(0)}(\x)= \left(1+\frac{\bnabla\circ\bnabla}{\om^2}\right) G_\om^{(0)}(\x),
\label{2.17}\ee
where $G_\om^{(0)}(\x)$ is the scalar Green's function
\be G_\om^{(0)}(\x)=\int \frac{d \k}{(2\pi)^3}\ \frac{e^{i\k\x}}{-\om^2+\k^2+i0}
=\frac{e^{i\om|\x|}}{4\pi|\x|}.
\label{2.18}\ee
Carrying out the derivatives in \Ref{3.8} we get
\be \G_\om^{(0)}(\x)=\left( \bm{A}-\frac{1-i\om|\x|}{(\om |\x|)^2}\bm{B}\right)
\ \frac{e^{i\om|\x|}}{4\pi|\x|}
\label{2.19}\ee
with
\be \bm{A}=1-\frac{\x\circ\x}{|\x|^2},\ \ \ \bm{B}=1-3\frac{\x\circ\x}{|\x|^2}.
\label{2.20}\ee
Eq. \Ref{2.19} is the well known retarded potential of a dipole.

The solution \Ref{2.15} for the electric field can be inserted into the second equation in \Ref{2.11}, which gives
\begin{widetext}
\be \sum_j
\left(m(-\om^2-i\ga\om+\Om^2)\delta_{ij}-4\pi e^2 \om^2 \G^{(0)}_\om(\a_i-a_j)\right)
\txi_{\om,j}=\tF_{\om,i}.
\label{2.21}\ee
\end{widetext}
This is an algebraic equation. Its solution can be written in the form
\be \txi_{\om,i}=\frac{1}{mN(\om)}\sum_j\L^{-1}_{ij}(\om)\tF_{\om,j},
\label{2.22}\ee
where $\L^{-1}_{ij}(\om)$ is the inverse of
\be \L_{ij}(\om)=\delta_{ij}-\al(\om)\om^2 \G^{(0)}_{\om}(\a_i-\a_j)
\label{2.23}\ee
and where we defined
\be N(\om)= -\om^2-i\ga\om+\Om^2, \quad \al(\om)=\frac{4\pi e^2}{mN(\om)}.
\label{2.24}\ee
Finally, we insert \Ref{2.22} into \Ref{2.15},
\be \tE_\om(\x)=\frac{\al(\om)\om^2}{e}\sum_{i,j}
\G^{(0)}_{\om}(\x-\a_i) \L^{-1}_{ij}(\om)\tF_{\om,j},
\label{2.25}\ee
and have with Eqs. \Ref{2.22} and \Ref{2.25} the solutions of the inhomogeneous equations \Ref{2.11} for the displacements and for the electric field, expressed in terms of the Langevin forces.

For the relation of the free energy with the vacuum energy, which will be discussed below, it is meaningful to consider  the first way too. We solve the second equation in \Ref{2.11},
\be \txi_{\om,i}  =  \frac{1}{mN(\om)}\left(e\tE_{\om}(\a_i)+\tF_{\om,i}\right),
\label{2.26}\ee
and insert the result into the first equation,
\begin{widetext}
\be  \left(-\om^2-\Delta+\bnabla\circ\bnabla
    -\al(\om)\om^2\sum_i\delta^{(3)}(\x-\a_i)    \right)\tE_\om(\x)
    =\frac{\al(\om)\om^2}{e}\sum_i\delta^{(3)}(\x-\a_i)\tF_{\om,i}\,.
\label{2.27}\ee
\end{widetext}
This is an effective equation for the electric field.  It is like a Schr{\"o}dinger equation with delta function potentials. This problem, especially equations like \Ref{2.27}, were recently discussed in \cite{bord15-91-065027}. The delta functions in this equation are three dimensional  and therefor the equation is ill defined. All known methods to handle this situation were  discussed in \cite{bord15-91-065027}. In terms of electrodynamics, one needs to exclude from the solution of Eq. \Ref{2.27} the action of the electric field created from a dipole acting on the dipole itself, i.e., its self-field. With these remarks, Eq. \Ref{2.27} can be easily solved. For that we introduce the corresponding Green's function, $\G_\om(\x,\x')$, obeying
\begin{widetext}
\be  \left(-\om^2-\Delta+\bnabla\circ\bnabla
    -\al(\om)\om^2\sum_i\delta^{(3)}(\x-\a_i)    \right)\G_\om(\x,\x')
    =\delta^{(3)}(\x-\x').
\label{2.28}\ee
\end{widetext}
This Green's function is related to the T-operator by
\bea \G_\om(\x,\x')&=&
\G_\om^{(0)}(\x-\x')
\label{2.29}\\&&
+\sum_{i,j}\G_\om^{(0)}(\x-\a_i)\T_{ij}(\om)\G_\om^{(0)}(\a_j-\x'),
\nn\eea
with the free Green's function defined in \Ref{2.16}. Inserting \Ref{2.29} into \Ref{2.28}, one obtains the equation,
\be \sum_k \left(\delta_{ij}-\al(\om)\om^2\G_\om^{(0)}(\a_i-\a_k)\right)\T_{kj}(\om)=
\al(\om)\om^2\delta_{ij},
\label{2.30}\ee
for the T-operator. Going back to eq. \Ref{2.23}, we note its relation to the $\L_{ij}(\om)$,
\be \T_{ij}(\om)= \al(\om)\om^2 \L_{ij}^{-1}(\om),
\label{2.31}\ee
%
which allows for the expression,
\bea \tE_{\om}(\x) &=&\frac{1}{e}\sum_{i,jj}\G^{(0)}_\om(\x,\a_i)\bm{T}_{ij}(\om)\tF_{\om,j},
\nn\\ \txi_{\om, i} &=& \frac{1}{4\pi e^2 \om^2}\sum_j\bm{T}_{ij}(\om)\tF_{\om,j},
\label{2.32}\eea
of the electric field and the displacements in terms of the T-operator.

These formulas will represent the correct solution only if we account for the remark on the self-field. It can be excluded by writing the  T-operator in the form
\be T_{ij}^{-1}(\om)=
\frac{1}{\al(\om)\om^2} \left( \delta_{ij}-\al(\om)\om^2\G_{\om, ij}^{(0)}\right),
\label{2.32a}\ee
where
\be \G_{\om,ij}^{(0)}  = \left\{\begin{array}{cl}0,&i=j,\\\G^{(0)}_\om(\a_i-\a_j),& i\ne j,\end{array}
\right.
\label{2.32b}\ee
i.e., where the diagonal terms were dropped. This way, Eq. \Ref{2.32} and \Ref{2.32a} describe the solutions for the electric field and for the displacements in case of dissipation.
\subsection{\label{T2.3}Thermal averages and the free energy}
In this subsection we calculate the internal energy $U$ by taking the thermal averages of the Hamiltonian \Ref{2.5}. We insert \Ref{2.32} and use the averages \Ref{2.12}. In these averages we have to insert the corresponding time dependent quantities, i.e., we have to go back using \Ref{2.10}. Now suppose, a field $A(t)$ has the Fourier representation
\be \hat{A}(t) =  \int_{-\infty}^\infty \frac{d\om }{2\pi} e^{-i\om t} \sum_i h_i(\om) \tF_{\om, i}.
\label{2.33}\ee
The corresponding thermal average, using \Ref{2.12}, is then
\begin{widetext}%
\bea  \langle \hat{A}(t)\hat{A}(t) \rangle
    &=&  \int_0^\infty\frac{d\om}{2\pi} \int_0^\infty\frac{d\om'}{2\pi}
    e^{i(\om+\om')t} \sum_{i,j} h_i(\om) h_j(\om')\langle \tF_{\om, i}\tF_{\om', j} \rangle,
\nn\\
&=& \int_0^\infty \frac{d\om}{\pi}\,  {\ga m \hbar \om} \coth\frac{\beta\om}{2}\sum_i h_i(\om)h_i(-\om).
\label{2.34}\eea
%
With these formulas, substituting
$h_i(\om)\to \frac{1}{e}\sum_i \G_\om^{(0)}(\x,\a_j)\bm{T}_{ji}(\om)$ for the electric field and
$\al_i(\om)\to \frac{1}{m\omp^2\om^2}\bm{T}_{ji}(\om)$ for the displacement field, we get
\be U=\int_0^\infty\frac{d\om}{\pi}\,\ga m \om    \coth\frac{\beta\om}{2}
\sum_i M_{ii},
\label{2.35}\ee
with
%
\bea M_{ij}   &=& {\rm tr} \sum_{k,l}\ \bm{T}_{ik}(\om)\left[
\frac{1}{8\pi e^2}
\int d\x\, \G^{(0)}_\om(\a_k,\x) \left(1+\frac{-\Delta+\bnabla\circ\bnabla}{\om^2}\right) \G^{(0)}_\om(\x,\a_l)
\right. \nn \\ && ~~~~~~~~~~~~~~~~~~~~~~~~~~~~~~~~\left.
+\frac{m}{2}\frac{\om^2+\Om^2}{(4\pi e^2\om^2)^2}\delta_{kl}
\right] \bm{T}_{lj}(-\om),
\label{2.36}\eea
\end{widetext}
and the trace is over the spatial structure. In the first term, which results from $E_{\rm ED}$, \Ref{2.6}, the second term in the parenthesis results from the magnetic field.


The first term in \Ref{2.36} can be simplified taking the Green's functions in momentum representation \Ref{2.17} and \Ref{2.18},
\bea &&\int d\x\, \G^{(0)}_\om(\a_k,\x) \left(1+\frac{-\Delta+\bnabla\circ\bnabla}{\om^2}\right) \G^{(0)}_\om(\x,\a_l)
\nn\\&&  =
\int \frac{d \k}{(2\pi)^3}\ {e^{i\k(\a_k-\a_l)}}m(k)
\nn\\&&=
\frac{1}{\om^4}\left(-1+\om\pa_\om\right)\om^2\G^{(0)}_\om(\a_k-\a_l),
\label{2.37}\eea
where we used the calculation,
\begin{widetext}
\be m(k) = \frac{1-\frac{\bm{k}\circ\bm{k}}{\om^2}}{-\om^2+k^2+i0}
\left(1+\frac{k^2-\bm{k}\circ\bm{k}}{\om^2}\right)
\frac{1-\frac{\bm{k}\circ\bm{k}}{\om^2}}{-\om^2+k^2+i0}
=
\frac{1}{\om^4}\left(-1+\om\pa_\om\right)\om^2\frac{1-\frac{\bm{k}\circ\bm{k}}{\om^2}}{-\om^2+k^2+i0},
\label{2.38}\ee
carried out in momentum space.
Inserting \Ref{2.37} into \Ref{2.36} delivers
\be M_{ij}=\frac{1}{2m(\omp^2\om^2)^2}\ {\rm tr} \sum_{k,l}  \bm{T}_{ik}(\om)\left((\om^2+\Om^2)\delta_{kl}
+\omp^2\left(-1+\om\pa_\om\right)\om^2 \G^{(0)}_{\om,ij}\right)\bm{T}_{lj}(-\om),
\label{2.39}\ee
\end{widetext}
where we used \Ref{2.32b} for excluding the self-field.
At this place it is meaningful to switch completely to matrix notations. We have to pay attention that we have two matrix structures, one structure resulting from the  dipoles, corresponding to the indices $i,j$,  and a (3x3) structure resulting from the spatial structures and that we must consider their tensor product.  Matrixes in the corresponding space are denoted by a hat. In this sense, Eq. \Ref{2.39} can be written in the form,
\begin{widetext}
\be \sum_i M_{ii}= \frac{m}{2 (4\pi e^2\om^2)^2} {\rm tr} \ \bm{\hat{T}}(\om)
\left(  \om^2+\Om^2+\omp^2 (-1+\om\pa_\om)\om^2 \bm{\hat{G}}^{(0)}_\om\right) \bm{\hat{T}}(-\om),
\label{2.40}\ee
%
where the trace is now over the whole space.

The last expression, \Ref{2.40}, can be rewritten in a more compact form. From \Ref{2.31} and \Ref{2.23}, we can write the inverse of the T-operator in the form
\be \frac{4\pi e^2}{m}\om^2\bm{\hat{T}}^{-1}(\om) =
-\om^2-i\ga\om+\Om^2- \frac{4\pi e^2}{m}\om^2\bm{\hat{G}}_\om^{(0)}
\label{2.41}\ee
and for its   derivative
\be \pa_\om\left( \frac{4\pi e^2}{m}\om^2 \bm{\hat{T}}^{-1}(\om)\right)=
-2\om-i\ga- \frac{4\pi e^2}{m}\pa_\om \left(\om^2\bm{\hat{G}}_\om^{(0)}\right)
\label{2.42}\ee
holds. We compose
%
\bea  && \pa_\om \left( \frac{4\pi e^2}{m}\om^2 \bm{\hat{T}}^{-1}(\om)\right)
                \left( \frac{4\pi e^2}{m}\om^2 \bm{\hat{T}}^{-1}(-\om)\right)
    -\left( \frac{4\pi e^2}{m}\om^2 \bm{\hat{T}}^{-1}(\om)\right)
            \pa_\om\left( \frac{4\pi e^2}{m}\om^2 \bm{\hat{T}}^{-1}(-\om)\right)
\nn\\  &&
    =-2i\ga\left[\om^2+\Om^2- \frac{4\pi e^2}{m}\left(-1+\om\pa_\om\right)
    \left(\om^2\bm{\hat{G}}_\om^{(0)}\right)\right].
\label{2.43}\eea
\end{widetext}
This relation allows to represent $\sum_iM_{ii}$, \Ref{2.40}, in the form
\be \sum_i M_{ii}=\frac{1}{ 2m\ga}\delta(\om)
\label{2.44}\ee
with
\be \delta(\om)=\frac{1}{2i}{\rm tr}\ln\frac{\bm{\hat{T}}(\om)}{\bm{\hat{T}}(-\om)}.
\label{2.45}\ee

This way, from \Ref{2.35} we get for the internal energy
\be U=\frac12\int_0^\infty \frac{d\om}{\pi}\, \om \coth\frac{\beta\om}{2}\, \pa_\om \delta(\om),
\label{2.46a}\ee
and, using \Ref{2.14}, for the free energy,
\be
F= \int_0^\infty \frac{d\om}{\pi}\, \left(\frac{\om}{2}+T\ln\left(1-e^{-\beta\om}\right) \right) \, \pa_\om \delta(\om).
\label{2.46}\ee
With Eqs. \Ref{2.46} and \Ref{2.45}, we have a representation of the free energy in terms of the T-operator related to the effective equation \Ref{2.28} for the electric field.

On a formal level (at least), it is possible to establish also a relation to the Green's function $\G_\om(\x,\x')$, defined by Eq. \Ref{2.28}. In matrix notation, we have from \Ref{2.28} as formal solution
\be \bm{\hat{G}}_\om
= \bm{\hat{G}}_\om^{(0)} \left(1-\bm{\hat{V}}\bm{\hat{G}}_\om^{(0)}\right)^{-1}
= \bm{\hat{G}}_\om^{(0)} \left(1+\bm{\hat{T}}(\om)\bm{\hat{G}}_\om^{(0)}\right),
\label{2.47}\ee
where the last equality follows from    \Ref{2.29}.
Written in matrix form, Eq. \Ref{2.30} reads
$\left(1-\bm{\hat{V}}\bm{\hat{G}}_\om^{(0)}\right)\bm{\hat{T}}= \bm{\hat{V}} $,
and can be rewritten in the form
\be \bm{\hat{V}}\left(1+\bm{\hat{G}}_\om^{(0)}\bm{\hat{T}}(\om) \right) = \bm{\hat{T}}.
\label{2.48}\ee
The potential $\bm{\hat{V}}$ is in our case $V(\x)=\al(\om)\om^2\sum_i\delta^{(3)}(\x-\a_i)$, which was the result of the interaction with the dipoles. Now, we use the chain of relations,
\bea {\rm tr}\ln\bm{\hat{T}}(\om)
&=&  {\rm tr}\ln \bm{\hat{V}}\left(1+\bm{\hat{G}}_\om^{(0)}\bm{\hat{T}}(\om)\right)
\nn\\&=&  {\rm tr}\ln \bm{\hat{G}}_\om -  {\rm tr}\ln \bm{\hat{G}}_\om^{(0)} +  {\rm tr}\ln \bm{\hat{V}}.
\label{2.52}\eea
On the formal level, we did not care about the existence of the individual contributions, especially about the ultraviolet divergences which are there.
Next, we drop the last two terms as giving a constant contribution to the free energy. This is justified, especially if having in mind as final goal the Casimir or Casimir-Polder forces. If doing so, we can write, in place of \Ref{2.45},
\be \delta(\om)=\frac{1}{2i} {\rm tr}\ln \frac{\bm{\hat{G}}_\om}{\bm{\hat{G}}_{-\om}}.
\label{2.53}\ee
Also, provided  $\bm{\hat{G}}_{-\om}=\bm{\hat{G}}_\om^*$ holds, a further rewriting delivers
\be \delta(\om)=\Im \ {\rm tr}  \ln  {\bm{\hat{G}}_\om} .
\label{2.54}\ee
Inserted into the free energy, \Ref{2.46}, the relation to the 'remarkable formula', mentioned in the Introduction, is more direct.
\subsection{\label{T2.4}The considered configurations}
We are going to apply the formulas developed above for a generic collection of dipoles, represented by their displacements $\txi_{\om, i}$. to three specific configurations, which we describe in this subsection.
These configurations are:
\ben
\item Atom-Atom (A-A)\\
Here we consider two atoms at positions $\a_i$ ($i=1,2$). Each atom is modeled as a dipole with dipole moment $\bm{p}_i(t)=e\bxi_i(t)$ and  has its own displacement vector $\bxi_i(t)$ ($i=1,2$). The corresponding current density is
\be \j(\x) = e\sum_{i=1}^2\pa_t\bxi_i(t)\delta^{(3)}(\x-\a_i).
\label{2.55}\ee
This is  basically the same equation as \Ref{2.3}, only the index $i$ is now restricted to $i=1,2$. In a similar way, we get for the equations \Ref{2.11},
\bea    (-\om^2-\Delta+\bnabla\circ\bnabla)\tE_\om(\x) &=&
4\pi e\om^2\sum_{i=1}^2\txi_{\om, i}\delta^{(3)}(\x-\a_i),
\nn\\
m(-\om^2-i\ga\om+\Om^2)\txi_{\om, i}  &=& e\tE_\om(\a_i)+\tF_{\om,i},\ (i=1,2).\nn\\
\label{2.56}\eea
At this place we mention, that, in general, the parameters of the atoms, for instance their intrinsic frequency $\Om$ or their mass, can be chosen individually for each. However, in order to keep formulas as simple as possible, we take these parameters equal for both atoms and do the same in the other configurations, too.
\item Atom-Wall (A-W)\\
For the atom we have a single dipole, which we give the number $i=0$, at location $\a_0$ with $a_z>0$, i.e., in the half space above the (x,y)-plane, and with displacement $\bxi_{0}(t)$.
%
%
We give its parameters entering \Ref{2.24} an index '0', i.e., we use $\ga_0$, $\Om_0$  and $\om_{p,0}$.
For the wall we assume a collection of dipoles with $i=1,2,\dots$, which are distributed homogeneously in the half space $z<0$, i.e., below the (x,y)-plane. Further, we increase their density to form a continuous distribution of dipoles, such that their displacements turn into a displacement field, $\bxi_{i}(t)\to\bxi(t,\x)$. Accordingly, the dipole moments turn into a continuous polarization, $\bm{p}_i(t)\to \bm{p}(t,\x)$. In all formulas for this transition we use
\bea \a_i&\to& \x, \ \ \sum_{i=1}^\infty \to \rho\int d\x \, \Theta(-z), \nn\\
\delta_{ij}&\to& \rho \delta^{(3)}(\x-\x'),
\label{2.57}\eea
where $\rho$ is the density (number per unit volume) of the dipoles in the wall.
In this configuration, the equations for the electric field and for the displacements are, again after Fourier transform in time,
\begin{widetext}
\bea    (-\om^2-\Delta+\bnabla\circ\bnabla)\tE_\om(\x) &=&
4\pi e\om^2\txi_{\om, 0}\delta^{(3)}(\x-\a_0)
+4\pi e\rho\, \om^2 \txi_\om(\x)\Theta(-z),
\nn\\
m(-\om^2-i\ga_0\om+\Om_0^2)\txi_{\om, 0}  &=& e\tE_\om(\a_0)+\tF_{\om,0},
\nn\\
m(-\om^2-i\ga\om+\Om^2)\txi_{\om}(x)  &=& e\tE_\om(\x)+\tF_{\om}(x), \ \ (z<0),
\label{2.58}\eea
\end{widetext}
which, in fact, constitute a special case of \Ref{2.11}. Again, we have a Langevin force, $\tF_{\om,0}$, for the atom and a continuum of these, $\tF_{\om}(\x)$, for the polarization of the medium.
\item Wall-Wall (W-W)\\
Here we assume two half spaces, one at $z<0$ and the other at $z>a$,  both filled with medium having the same properties as in the A-W configuration. The displacement field, $\txi_\om(\x)$, exists in the two half spaces, $z<0$ and $z>a$. The gap between the half spaces has width $a$ and is assumed to be empty. The equation \Ref{2.11} take here the form,
\begin{widetext}
\bea    (-\om^2-\Delta+\bnabla\circ\bnabla)\tE_\om(\x) &=&
4\pi e\rho\, \om^2 \txi_\om(\x)(\Theta(-z)+\Theta(z-a)),
\nn\\
m(-\om^2-i\ga\om+\Om^2)\txi_{\om}(x)  &=& e\tE_\om(\x)+\tF_{\om}(x), \ \ (z<0 \ \mbox{and} \ z>a),
\label{2.59}\eea
\end{widetext}
and $\tF_{\om}(x)$ is the continuum of the Langevin forces.
\een
The first two configurations, A-A and A-W, correspond to the Casimir-Polder force \cite{casi48-73-360}, and the third one, W-W, to the Casimir effect, \cite{casi48-51-793} for ideal conductors and \cite{lifs56-2-73} for dielectric slabs.
\section{\label{T3}Atom-Atom configuration}
In this section we investigate in detail the A-A configuration, i.e., the interaction of two dipoles at separation $a=|\a_1-\a_2|$, as introduced in Sec. \ref{T2.4}. The relevant equations are \Ref{2.56}. The calculation of the free energy goes literally the same way as in Sec. \ref{T2.3}, with the restriction of the indices to the values $i=1,2$. This way, the free energy is given by\Ref{2.46} and \Ref{2.45}, together with \Ref{2.31} and \Ref{2.23}. With these formulas we express $\delta(\om)$ in the form
\be \delta(\om)  =
 \frac{-1}{2i} {\rm tr}\ln \frac{\bm{\hat{L}}(\om)}{\bm{\hat{L}}(-\om)},
\label{3.1}\ee
which is rewritten as compared to \Ref{2.45} using \Ref{2.31}, dropping contributions, which do not depend on the separation between the atoms. The matrix ${\bm{\hat{L}}(\om)}$ is now (2x2) in the indices $i,j$ and, using \Ref{2.32a}, it can be rewritten in the form
\be {\bm{\hat{L}}(\om)}=1-\al(\om)\om^2\bm{\sigma}_1\G^{(0)}_\om,
\label{3.2}\ee
where with \Ref{2.32b} we defined $\G^{(0)}_\om=\G^{(0)}_{\om,12}$ (i.e., exclu\-ding the self-field) and $\bm{\sigma}_1=\left(\begin{array}{c}0,1\\1,0\end{array}\right)$.

The trace in \Ref{3.1} can be calculated in two steps. First, we consider the (2x2) structure related to the two atoms. It gives simply
\be {\rm tr}\ln {\bm{\hat{L}}(\om)} =
\sum_{\sigma=\pm 1}{\rm tr}\ln\left(1-\sigma\al(\om)\om^2\G^{(0)}_\om\right).
\label{3.3}\ee
The terms with $\sigma=\pm 1$ correspond to symmetric and antisymmetric solutions of Eq. \Ref{2.27}, which is a result of the symmetry under exchanging the two dipoles.
In \Ref{3.3}, the remaining trace is over the spatial (3x3)-structure. Now, $\G^{(0)}_\om$ is given by \Ref{2.18} with $\x=\a_1-\a_2$ and we can use the explicit formula \Ref{2.19} with $|\x|=a$. For the trace we use the formula
\be {\rm tr}\ln(1+p\bm{A}+q\bm{B})=2\ln(1+p+q)+\ln(1-2q),
\label{3.4}\ee
where $p$ and $q$ are some numbers, and  $\bm{A}$ and $\bm{B}$ are the matrices given by \Ref{2.20}. This way, and carrying out the sum over $\sigma=\pm1$,  we get
\be \delta(\om)=2\delta_1(\om)+\delta_2(\om)
\label{3.5}\ee
with
\be \delta_i(\om)=\frac{-1}{2i}\ln\frac{L_i(\om)}{L_i(-\om)},
\label{3.6}\ee
where we defined
\bea
L_1(\om) &=& 1-\left[ \frac{\al(\om)e^{i\om a}}{4\pi a^3}(1-i\om a-(\om a)^2)\right]^2,
\nn\\
L_2(\om) &=& 1-\left[ \frac{2\al(\om)e^{i\om a}}{4\pi a^3}(1-i\om a)\right]^2.
\label{3.7}\eea
With these formulas, and  inserting representation \Ref{3.5} for $\delta(\om)$ into $F$, \Ref{2.46}, we have explicit formulas for the free interaction energy of two atoms in case of dissipation. It should be remarked, that these expressions constitute a nonperturbative generalization of the Casimir-Polder interaction potential, which can be obtained backwards by expanding in powers of the coupling $\al(\om)$.

In the following, we consider limiting cases of the free energy and the transition to the Matsubara representation.

\subsection{\label{T3.1}Limiting cases}
In this subsection we consider the limiting cases of vanishing dissipation parameter and the relation to the vacuum energy and check the low temperature expansion.

\paragraph{Vanishing dissipation parameter and vacuum energy} We split the free energy \Ref{2.46} according to the phase, \Ref{3.5},
\be F=2F_1+F_2,\
\label{3.8a}\ee
with
\be
F_i= \int_0^\infty \frac{d\om}{\pi}\,
\left(\frac{\om}{2}+T\ln\left(1-e^{-\beta\om}\right) \right)
\,\sum_{\sigma=\pm1} \pa_\om \delta_i(\om).
\label{3.8}\ee
The $\delta_i(\om)$ are given by \Ref{3.6}. Instead, we use
\be \delta_i(\om)=-\frac{1}{2i}\ln\frac{\Phi_i(\om)-i\ga\om}{\Phi_i(\om)+i\ga\om},
\label{3.9}\ee
where
\bea
\Phi_1(\om)&=&-\om^2+\Om^2-\sigma\frac{e^2}{m a^3}e^{i\om a}, \nn \\
\Phi_2(\om)&=&-\om^2+\Om^2-2\sigma\frac{e^2}{m a^3}e^{i\om a},
\label{3.10}\eea
and, again using \Ref{2.24}, we dropped a factor, resulting in a contribution to the free energy which does not depend on the separation.

Next we consider the limit $\ga\to0$ and apply the Sokhotski-Plemelj theorem to
\be \pa_\om \delta_i(\om) = -\frac{1}{2i}\left(
 \frac{\pa_\om \Phi_i(\om)-i\ga } {\Phi_i(\om)-i\ga\om}
-\frac{\pa_\om \Phi_i(\om)+i\ga } {\Phi_i(\om)+i\ga\om} \right),
\label{3.11}\ee
which results in
\be \lim_{\ga\to 0} \pa_\om \delta_i(\om) =
-\pi\delta(\Phi_i(\om))\, \pa_\om (\Phi_i(\om)).
\label{3.12}\ee
Inserting into \Ref{3.8} gives for the free energy at vanishing dissipation
\begin{widetext}
\be {F_i}_{|_{\ga=0}} = -\int_0^\infty {d\om}\,
\left(\frac{\om}{2}+T\ln\left(1-e^{-\beta\om}\right) \right)
\delta(\Phi_i(\om))\, \pa_\om (\Phi_i(\om)).
\label{3.13}\ee
\end{widetext}
Using the delta function, the integration can be carried out. Let $\om_J$ be the zeros of the equation $\Phi_i(\om)=0$, i.e., the eigenfrequencies of the system of two dipoles interacting with the electromagnetic field without dissipation. Then the integral can be written as a sum,
\be  {F}_{|_{\ga=0}} = \sum_J\,
\left(\frac{\om_J}{2}+T\ln\left(1-e^{-\beta\om_J}\right) \right).
\label{3.14}\ee
Here we assumed the sum over $\sigma$ and over the two contributions $2F_1+F_2$ being included into the sum over $J$ and we accounted for $\pa_\om (\Phi_i(\om))<0$, which can be seen from \Ref{3.10} at least for small $\omp^2$.
The vacuum energy can now be obtained from \Ref{3.14} simply taking $T\to0$,
\be E_0= \left({F}_{|_{\ga=0}}\right)_{|_{T=0}} = \frac12 \sum_s\,\om_s,
\label{3.15}\ee
in agreement with \Ref{3}, again ignoring the necessary regularization.

At this place a further remark concerning the sum over the eigenfrequencies $\om_s$ is in order. As in detail discussed in \cite{D17-1}, the considered system of two dipoles does not have real eigenfrequencies in empty space, since any excitation would be radiated away. The way out is to put the system, including the electromagnetic field, into a box. Then the eigenfrequencies are real and the sum in \Ref{3.15} makes sense. Afterwards one may tend the volume to infinity  and obtains the vacuum energy \Ref{3.15}.

Another way to get the vacuum energy out from the free energy is to let $\ga\to0$  directly in \Ref{3.8}. Thereby one has to pay attention that the singularities of ${\Phi_i(\om)-i\ga\om}$ are in the lower half plane and from $-i\ga$ one has to keep an '$+i0$'-prescription. This can be done  in the following way,
\begin{widetext}
\be E_0= \left({F}_{|_{\ga=0}}\right)_{|_{T=0}} = -\frac12\int_0^\infty\frac{d\om}{\pi}\,\om\,
\pa_\om \frac{1}{2i} {\rm tr}
\left(2\ln \frac{L_1(\om-i0)}{L_1(\om+i0)}+\ln \frac{L_2(\om-i0)}{L_2(\om+i0)}\right)
\label{3.16}\ee
\end{widetext}
In this representation, it makes sense to integrate by parts and to make the Wick rotation. One arrives at
\be E_0=\frac12\int_0^\infty\frac{d\xi}{\pi}\,
\left(2\ln  {L_1(i\xi)} +\ln  {L_2(i\xi)} \right),
\label{3.17}\ee
with, explicitly,
\bea
L_1(i\xi) &=& 1-\left[ \frac{\al(i\xi)e^{- a\xi}}{4\pi a^3}\left(1+a\xi+(a\xi)^2\right)\right]^2,
\nn\\
L_2(i\xi) &=& 1-\left[ \frac{2\al(i\xi)e^{-a\xi}}{4\pi a^3}(1+a\xi)\right]^2.
\label{3.18}\eea
with $\al(i\xi)=\frac{\omp^2}{\xi^2+\Om^2}$.
These formulas constitute the  non-perturbative formulation of the Casimir-Polder interaction energy of two dipoles (without dissipation). We mention that the functions $L_i(i\xi)$ ($i=1,2$), may have zeros and change sign. This is the criticality mentioned in the Introduction. It happens with $L_2$ for $\al/(4\pi a^3)<1/4$ and with $L_1$ for $\al/(4\pi a^3)<1$ for constant $\al$ and for $e^2/\Om^2$ in place of $\al$ for $\al(\om)$ given by \Ref{2.24}, provided $\Om\ne0$. In case $\Om=0$, there is criticality for all values of the parameters.
This instability was mentioned in the literature earlier, for example in \cite{berm14-89-022127}, who also mentioned that the onset of criticality coincides with the radius of convergence of the long separation expansion.

At this place it is easy the check that \Ref{3.18} gives the correct long separation expansion. Expanding \Ref{3.18} in powers of $\al$, and substituting $\xi\to\xi/a$ we get,
\be E_0\cong - \frac{1}{a}\int_{0}^\infty \frac{d\xi}{\pi} \
    \left(\frac{\al(i \xi)}{4\pi a^3}\right)^2 (3+6\xi+5\xi^2+2\xi^3+\xi^4)\,e^{-2\xi},
\label{3.19}\ee
and carrying out the integration for constant $\al$, $E_0\cong-\frac{23}{4\pi a^7}\left(\frac{\al}{4\pi}\right)^2$ holds in agreement with Eq. (57) in \cite{casi48-73-360}.

\paragraph{Low temperature expansion} For the expansion for $T\to0$ of the temperature dependent part of the free energy,
\be \Delta_TF= T\int_0^\infty \frac{d\om}{\pi}\, \ln\left(1-e^{-\beta\om}\right)   \, \pa_\om \delta(\om),
\label{3.19a}\ee
which results from the second term in the parenthesis in \Ref{2.46},
we need the expansions of the phases $\delta_i(\om)$  for $\om\to0$. From \Ref{3.7} we get with \Ref{2.24} for $\Om\ne0$
\bea L_1(\om) &=& 1-\left(
    \frac{e^2 }{m a^3 \Om^2}\right)^2 +O(\om),
\nn\\
L_2(\om) &=& 1-
    \left(\frac{2e^2 }{m a^3 \Om^2}\right)^2+O(\om),
\label{3.20}\eea
and for the phase
\bea  \delta(\om)&=&\frac{6(\ga+2a\Om^2)\left(8\pi^2a^6\Om^4-\left(\frac{4\pi e^2}{m}\right)^2\right)}
    {\left(\omp^8 - 20 \pi^2\left(\frac{4\pi e^2}{m}\right)^2\Om^4+64\pi^4 a^{12}\Om^8 \right)\Om^2}\,\om
  \nn\\&&  +O(\om^3)
\label{3.21}\eea
holds. We note that formulas \Ref{3.20} do not depend on the dissipation parameter $\ga$. In \Ref{3.21} one may put $\ga=0$ without principal changes.

If we put $\Om=0$, these expressions simplify  and the $L_i$ become more singular,
\bea L_1(\om) &=& \left(
    \frac{e^2 }{m a^3\ga }\right)^2\frac{1}{\om^2} +O\left(\frac{1}{\om}\right),
\nn\\ L_2(\om) &=& \left(
    \frac{2e^2 }{m a^3 \ga}\right)^2\frac{1}{\om^2} +O\left(\frac{1}{\om}\right),
\label{3.22a}\eea
and for the phase,
\be  \delta(\om)=-\frac{6(1+2\ga a)}
    {\ga}\,\om+O(\om^3)
\label{3.22b}\ee
holds.
Finally, if we put in addition also $\ga=0$, we get an even more singular behavior,
\bea L_1(\om) &=& -\left(
    \frac{e^2 }{m a^3 }\right)^2\frac{1}{\om^4} +O\left(\frac{1}{\om^3}\right),
\nn\\
L_2(\om) &=& \left(
    \frac{2e^2 }{m a^3 }\right)^2\frac{1}{\om^4} +O\left(\frac{1}{\om^3}\right),
\label{3.22c}\eea
and for the phase
\be  \delta(\om)=-12a  \om+O(\om^3)
\label{3.23}\ee
holds. Below we need for $\Om\ne0$ the values $L_i(0)$, which are finite, and for $\Om=0$ the derivatives, which are for $\ga\ne0$
\be \pa_\om \ln L_i(\om)=-\frac{2}{\om}+O(1),
\label{3.24}\ee
and, in addition,  for $\ga=0$
\be \pa_\om \ln L_i(\om)=-\frac{4}{\om}+O(1).
\label{3.25}\ee
It must be mentioned that the leading orders of the expansions of the phases for the cases considered result from the next order contributions in the expansions of $L_i(\om)$, which are not shown. Also we mention that the phases $\delta(\om)$ always start from the first order in $\om$ since any constant contribution disappears under the derivative, $\pa_\om$,  in \Ref{3.19a}.

Summarizing, introducing the notation $c_1$ for the coefficient in the expansion \Ref{3.21}, \Ref{3.22b}, \Ref{3.23},
\be\delta(\om)=c_1\om+O(\om^3),
\label{3.25a}\ee
we get in the temperature dependent part of the free energy \Ref{3.19a}, after the substitution $\om\to\om T$,
\be \Delta_TF=T^2\int_0^\infty \frac{d\om}{\pi} \ \ln\left(1-e^{-\hbar\om}\right)
        \pa_\om \left(c_1\om+O(T^2\om^3)\right),
\label{3.27}\ee
which gives
\be \Delta_TF=-\frac{c_1\zeta(2)}{\pi\hbar}T^2 +O(T^4),
\label{3.28}\ee
where $\zeta(2)=\pi^2/6$ is the Riemann zeta function.
In this formula, $c_1$ is first non zero coefficient in the expansion of the phase for $\om\to0$ for all values of the parameters. Thus, the low temperature behavior of the entropy \Ref{2.16},
\be S=2c_1 T+O(T^2),
\label{3.29}\ee
does never violate the third law of thermodynamics.
We mention that the same holds if we allow for a temperature dependence of the dissipation parameter, $\ga(T)$, with a low-T behavior
$\ga(T)\raisebox{-4pt}{${ \sim \atop T\to0}$} \ga_1 T^\al$ ($\al>1$). With $\Om=0$ we get after the substitution $\om\to\om T$
\be \al(\om T)=\frac{4\pi e^2}{m\left(-\om^2T^2-i\ga_1 T^{\al+1}\right)}.
\label{3.29a}\ee
Now,   for $\al<1$ this is equivalent to the case \Ref{3.22a} and \Ref{3.22b}, and for $\al>1$ it is equivalent to \Ref{3.22c} and \Ref{3.23}. This way, in the A-A configuration, also the temperature dependence of the dissipation parameter does not result in thermodynamic problems.

\subsection{\label{T3.2}Transition to Matsubara frequencies}
Starting point for the transition to Matsubara frequencies is the free energy
\begin{widetext}
\be F=\int_0^\infty\frac{d\om}{\pi}\,
\left[\frac{\om}{2}+T\ln\left(1-e^{-\beta \om}\right)\right]
\pa_\om\frac{-1}{2i}
\left(2\ln\frac{L_1(\om)}{L_1(-\om)}+\ln\frac{L_2(\om)}{L_2(-\om)}
\right),
\label{3.30}\ee
\end{widetext}
which follows from  \Ref{3.19a}, \Ref{3.5} and \Ref{3.7}. We mention that $L_i(\om)$ has no singularities in the upper half plane (at least for sufficiently small $\ga$). Thus we write the logarithms of the quotients as differences of two logarithms and turn the integration path upwards towards the imaginary axis in the first term and in the second term correspondingly downwards, which is in fact a Wick rotation.

Below we will need the following simple relation,
\be \frac{\beta\om}{2}+\ln\left(1-e^{-\beta\om}\right)=\ln\left(2\sinh\frac{\beta\om}{2}\right)
\label{3.31}\ee
and its continuation,
\be \ln\left(2\sinh\frac{i\beta\xi}{2}\right)
=\ln\left| 2\sin\frac{\beta\xi}{2}\right |
+i\pi \sump \Theta(\xi-\xi_l),
\label{3.32}\ee
where $\xi_l=2\pi T l$ are the Matsubara frequencies. This relation follows since the logarithm has cuts starting in $\xi=\xi_l$. The prime on the sum symbol denotes, as usual, that the contribution from $l=0$ must be taken with a factor $1/2$.

However, in doing the Wick rotation, we must split the integral over the difference   into the difference of two integrals. In doing so,  the integration over $\om$ will not converge at $\om=0$
in the cases  with $\Om=0$ because of the singular behavior in  \Ref{3.22a} and \Ref{3.22c}. We note, that this is not the case for $\Om\ne0$, as can be seen from \Ref{3.20}.

The difficulty arising at $\om=0$ can be handled  introducing a lower integration boundary $\ep>0$ and taking  the limit $\ep\to0$ after the rotations. This way we get
\begin{widetext}
%
\be F_i  =
-\frac{1}{2\pi i}\lim_{\ep\to0}
     \left[\int_\ep^\infty d\om\, \left(\frac{\hbar\om}{2}+T\ln\left(1-e^{-\beta\hbar\om}\right)\right) \pa_\om \ln L_i(\om)
- \int_\ep^\infty d\om\,\left(\frac{\hbar\om}{2}+T\ln\left(1-e^{-\beta\hbar\om}\right)\right) \pa_\om \ln L_i(-\om)\right],
\label{3.33}\ee
\end{widetext}
where we have split the free energy, $F=2F_1+F_2$, in accordance with the split \Ref{3.5} of the phase.

Now we split the integration pathes into two parts each. The first part is a half circle of radius $\ep$, given by $\om=\ep e^{i\varphi}$ in the first integral and $\om=\ep e^{-i\varphi}$ in the second one,
with $\varphi=0 \dots \frac{\pi}{2}$ in both cases. The second part is a straight line along the imaginary axis with  $\om=i\xi$  in the first integral and $\om=-i\xi$  in the second one, with $\xi\in [\ep,\infty)$ in both cases.
From the integrals along the straight lines, using \Ref{3.32},  we have
\begin{widetext}
\bea F_i^{\rm lin.}&=&-\frac{T}{2\pi i} \left[ \int_\ep^\infty d\xi\,
    \left( \ln \left|2\sin\frac{\beta\xi}{2}\right|
-i\pi\sump\Theta(\xi-\xi_l)\right)
            \pa_\xi\left(\ln\left|L_i(i\xi)\right|+i\pi\Theta(\xi_*-\xi)\right)
     \right.\nn\\&& \left.~~~~
   -\int_\ep^\infty d\xi\,  \left( \ln \left|2\sin\frac{\beta\xi}{2}\right|
+i\pi\sump\Theta(\xi-\xi_l)\right)
            \pa_\xi\left(\ln\left|L_i(-i\xi)\right|-i\pi\Theta(\xi_*-\xi)\right) \right].
\label{3.34}\eea
\end{widetext}
An additional feature appeared here in case the functions $L_i(-i\xi)$ have a zero, i.e., there is a real $\xi_*>0$ such that $L_i(-i\xi_*)=0$ holds, as discussed in Sect \ref{T3.1}. For this reason, the logarithms of $L_i$ acquire imaginary parts as shown in the above formula. The signs follow from $L_i(i\xi)>0$ for sufficiently large $\xi$ and the side on which the branch point of the logarithm is passed.

After simplifying in Eq.\Ref{3.34}, we integrate by parts. In the contribution with $l=0$,   accounting for $\Theta(\xi-\xi_0)=1$ for $\xi>0$, we get
$\frac{T}{2} \ln\left|L(i\ep)\right|$. In all other contribution we can put $\ep=0$. This way, from \Ref{3.34}
\be F_i^{\rm lin.}=\frac{T}{2} \ln\left|L_i(i\ep)\right|+T\sum_{l=1}^\infty \ln\left|L_i(i\xi_l)\right|
  -T \ln\left|2\sin\frac{\beta\xi_*}{2}\right|
\label{3.35}\ee
follows.
Indeed, this expression is divergent for $\ep\to0$ in case of $\Om=0$. For $\Om\ne0$, Eqs. \Ref{3.20} show finite $L_i(0)$. For $\Om=0$, using \Ref{3.22a}, we get for $\ga\ne0$,
\be \frac{T}{2} \ln\left|L_1(i\ep)\right|
    =-T\ln\ep
    +T\ln\left(\frac{ e^2}{m a^3\ga}\right)+O(\ep),
\label{3.36}\ee
and for $\ga=0$, using \Ref{3.22c},
\be \frac{T}{2} \ln\left|L_1(i\ep)\right|
    =-2T\ln\ep
    +T\ln\left(\frac{ e^2}{m a^3}\right)+O(\ep).
\label{3.37}\ee
The corresponding expressions for $L_2(i\ep)$ can be obtained from these by the substitution $\omp^2\to 2\omp^2$. We see that in these cases, i.e., for $\Om=0$,  there is a logarithmic divergence.

It remains to calculate the contribution from the half circles.  Restricting to contributions not vanishing for $\ep\to0$  we note
\be \frac{\hbar\om}{2}+T\ln\left(1-e^{-\beta\hbar\om}\right)\sim
    T\ln(\beta\ep e^{i\varphi})
\nn\ee
in the first contribution in \Ref{3.33} and the complex conjugate of that in the second. Using the leading contribution in \Ref{3.24} or \Ref{3.25}, we get from \Ref{3.33}
\bea F_i^{\rm h.circ.} &=& -\frac{T}{\pi}\int_0^{\pi/2} d\varphi\,
    \left(\ln(\beta\ep e^{i\varphi})+\ln(\beta\ep e^{-i\varphi})\right),
\nn\\&=&   T\ln(\beta\ep)+O(\ep).
\label{3.39}\eea
for $\ga\ne0$ and
\bea F_i^{\rm h.circ.} &=& -\frac{T}{\pi}\int_0^{\pi/2} d\varphi\,
    \left(\ln(\beta\ep e^{i\varphi})+\ln(\beta\ep e^{-i\varphi})\right),
\nn\\&=&  2 T\ln(\beta\ep)+O(\ep)
\label{3.40}\eea
for $\ga=0$.
We see, the divergence for $\ep\to0$ compensates just that in \Ref{3.36} or \Ref{3.37}, as expected, and we get for the sum of \Ref{3.35} and \Ref{3.40}, for $\ga\ne0$,
\begin{widetext}
\be F_i=
T\ln\left(\frac{c_i \omp^2\beta}{4\pi a^3\ga}\right)
    +T\sum_{l=1}^\infty \ln\left|L_1(i\xi_l)\right|
    -T \ln\left|2\sin\frac{\beta\xi_*}{2}\right|,\quad (i=1,2),
\label{3.41}\ee
with $c_1=1$, $c_2=2$, and from \Ref{3.35} and \Ref{3.37}, for $\ga=0$, we get
\be F_i=
2T\ln\left(\frac{c_i \omp^2\beta}{4\pi a^3}\right)
    +T\sum_{l=1}^\infty \ln\left|L_1(i\xi_l)\right|
    -T \ln\left|2\sin\frac{\beta\xi_*}{2}\right|,\quad (i=1,2){\color{red} . }
\label{3.42}\ee
\end{widetext}
{\color{red}W}ith these formulas we derived the Matsubara representation of the complete free energy $F=2F_1+F{\color{red}{_2}}$ for the non-perturbative Casimir-Polder interaction of two dipoles with dissipation in case of vanishing intrinsic frequency $\Om=0$. For $\Om\ne0$, the same formulas hold with the usual terms corresponding to zeroth Matsubara frequency $L_i(i\xi_{|_{l=0}})$ in place of the first logarithmic contributions. In other words, the modifications, coming in for $\Om=0$, can be formulated in terms of the formal substitutions of the contributions from the zeroth Matsubara frequency according to, for $\ga\ne0$,
\be  L_1(i\xi_{|_{l=0}}) \to
\left(\frac{ e^2\beta}{m a^3\ga}\right)^2,
\quad
L_2(i\xi_{|_{l=0}}) \to
\left(\frac{ 2e^2\beta}{m a^3\ga}\right)^2,
\label{3.43}\ee
and, for $\ga=0$,
\be  L_1(i\xi_{|_{l=0}}) \to
\left(\frac{ e^2\beta}{m a^3 }\right)^4,
\quad
L_2(i\xi_{|_{l=0}}) \to
\left(\frac{ 2e^2\beta}{m a^3 }\right)^4.
\label{3.44}\ee
The other modification in \Ref{3.41} and \Ref{3.42}, i.e., the second logarithmic contributions,  comes from the critical behavior, which may be present for $\Om\ne0$ and will be present for $\Om=0$. We mention that for $\Om=0$ the contributions from the zeroth Matsubara frequency has, for $T\to0$, a logarithmic behavior, $F\sim - T \ln T$.

\section{\label{T4}Atom-Wall configuration}
In this section we investigate in detail the A-W configuration, i.e., the interaction of a single dipole with a dielectric half space, as introduced in Sec. \ref{T2.4}. As before, the free energy is given by Eq. \Ref{2.46}. The phase $\delta(\om)$ is given, in general notation, by \Ref{2.45}. Using \Ref{2.31}, we express $\delta(\om)$ through $\bm{\hat{L}}(\om)$,
\be \delta(\om)=-\frac{1}{2i}{\rm tr}\ln \frac{\bm{\hat{L}}(\om)}{\bm{\hat{L}}(\om)^*},
\label{4.0}\ee
dropping terms which would not contribute to the Casimir-Polder force. In this formulas we also used $\bm{\hat{L}}(-\om)=\bm{\hat{L}}(\om)^*$. With \Ref{2.31} and \Ref{2.32a} we have
\be \L_{ij}(\om)=\delta_{ij}-\al(\om)\om^2 \G^{(0)}_{\om,ij}
\label{4.1}\ee
for the matrix elements of $\bm{\hat{L}}(\om)$.

For the A-W configuration, the index $i=0$ describes the atom and the indices $i=1,2,\dots$, describe the dipoles in the half space $z<0$. The separation between the atom and the wall is $a=(\a_0)_z$.

For the subsequent derivations we return for a moment to Eq. \Ref{2.28} for the Green's function,
\be  \left(-\om^2-\Delta+\bnabla\circ\bnabla
    -V(\x)   \right)\G_\om(\x,\x')
    =\delta^{(3)}(\x-\x'),
\label{4.2}\ee
where we introduced the notation
\be V(\x)=V_s(\x)+V_1(\x),\label{4.3v}\ee
with
\bea
V_s(\x) &=& \al(\om)\om^2 \delta^{(3)}(\x-\a_0),\nn \\
V_1(\x) &=& \al(\om)\om^2\sum_{i=1}^\infty\delta^{(3)}(\x-\a_i).
\label{4.3}\eea
Here, $V_s(\x)$ is the potential from the single atom with $\al(\om)$ given by Eq. \Ref{2.24}, and $V_1(\x)$ is the potential from the dipoles in the half space.
Next, we switch to operator and matrix notations as used already at the end of Sec. \ref{T2.3}. From Eqs. \Ref{4.1} and \Ref{4.3} we have
\be {\rm tr}\ln \bm{\hat{L}}(\om)= {\rm tr}\ln\left( 1-(\bm{\hat{V}}_s+ \bm{\hat{V}}_1 ) \hG^{(0)}_\om\right).
\label{4.4}\ee
We continue by rewriting
\bea  {\rm tr}\ln \bm{\hat{L}}(\om)&=& {\rm tr}\ln (1-\bm{\hat{V}}_1 \hG^{(0)}_\om )
 \label{4.5}\\&&   + {\rm tr}\ln \left(1-(1-\bm{\hat{V}}_1\hG^{(0)}_\om )^{-1} \bm{\hat{V}}_s\hG^{(0)}_\om\right) ,
\nn\eea
where the first term does not depend on the atom-wall separation and will be dropped  as not contributing to the Casimir-Polder force. In the other term, we use the cyclic property of the trace and an equation which is analog to \Ref{2.47} with $V_1(\x)$ in place of $V(\x)$, defining the Green's function $\hG^{(1)}_\om(\x,\x')$, which is the Green's function in the presence of the dipoles in the half space alone, and come to
\be  {\rm tr}\ln \bm\hat{{L}}(\om) =  {\rm tr}\ln (1-\bm{\hat{V}}_s \hG^{(1)}_\om )+\dots\,,
\label{4.6}\ee
which  we will use in the following. The next step is the limiting transition $\a_i\to\x$ with $i=1,2,\dots$ to a continuous medium in $z<0$. Thereby the potential $V_1(\x)$ turns into
\be V_1(\x)=(\ep(\om)-1)\om^2\Theta(-z)
\label{4.3b}\ee
and we introduced the permittivity
\be \ep(\om)=1+\frac{\omp^2}{N(\om)},\ \ \omp^2=\frac{4\pi e^2}{m}\rho,
\label{4.3a}\ee
where $\rho$ is the density (number of atoms per unit volume). Representing
\be \hG^{(1)}_\om  =  \hG^{(0)}_\om + \Delta \hG^{(1)}_\om
\label{4.7}\ee
as a sum of the free space Green's function \Ref{2.17} and an addendum from the  half space, we note
\be \G^{(1)}_\om(\x,\x') = \lim_{\a_i\to \x \atop \a_j\to \x'}
    \left(\G^{(0)}_\om (\a_i-\a_j)+\Delta \G^{(1)}_\om(\a_i-\a_j) \right).
\label{4.8}\ee
Now, in \Ref{4.6}, the arguments $\x$ and $\x'$ will be put equal one another due to the potential $V_s(\x)$, \Ref{4.3}. Thus, the free Green's function $\G^{(0)}_\om (\a_i-\a_j)$ must be set equal to zero in order to avoid self-fields, and we are left with the addendum and come to
\be  {\rm tr}\ln \bm\hat{{L}}(\om) =  {\rm tr}\ln
\left(1-\bm{\hat{V}}_s \, \Delta \hG^{(1)}_\om\right)+\dots\,.
\label{4.9}\ee
After performing the transition to the continuum, the Green's function $\G^{(1)}_\om(\x,\x')$ obeys the equation
\begin{widetext}
\be  \left(-\om^2-\Delta+\bnabla\circ\bnabla
    - (\ep(\om)-1)\om^2 \Theta(-z)   \right)\G^{(1)}_\om(\x,\x')
    =\delta^{(3)}(\x-\x') .
\label{4.10}\ee
%
Here, in view of the translational invariance in the (x,y)-plane, we perform the corresponding Fourier transform,
\be \G^{(1)}_\om(\x,\x') = \int\frac{dk_{||}}{(2\pi)^2}
e^{i\k_{||}(\x_{||}-\x'_{||})} \bm{{g}}_\Gamma(z,z')
\label{4.11}\ee
with $\Gamma=\sqrt{\om^2-k_{||}^2+i0}$ and the equation
%
\be  \left(-\Gamma^2-\pa_z^2+\bnabla\circ\bnabla
    - (\ep(\om)-1)\om^2 \Theta(-z)   \right)
    \bm{{g}}_\Gamma(z,z')=\delta^{(1)}(z-z')
\label{4.12}\ee
\end{widetext}
holds. In this representation, the gradient is $\bnabla=(i\k_{||},\pa_z)$. This Green's function is well known, see for example, App. A in \cite{sche08-58}, or Chap. 13 in \cite{milton98}. We use its decomposition into polarizations,
\be \bm{{g}}_\Gamma(z,z') =
\E_{\rm TE} \ {{g}}_\Gamma^{\rm TE}(z,z') \ \E_{\rm TE}^\dagger
+ \E_{\rm TM}\ {{g}}_\Gamma^{\rm TM}(z,z') \ \E_{\rm TM}^\dagger,
\label{4.13}\ee
with the polarization vectors
\be \E_{\rm TE} =\left(\begin{array}{c}-k_2 \\ k_1 \\ 0\end{array}\right) \frac{1}{k_{||}},
\quad \E_{\rm TM} =\left(\begin{array}{c}-ik_1 \pa_z\\ ik_2\pa_z \\ -k_{||}^2\end{array}\right) \frac{1}{k_{||}\om},
\label{4.14}\ee
and we dropped the delta function contribution at $z=z'$ since it results from the free space part in \Ref{4.7}, which we drop.

For the scalar functions ${{g}}_\Gamma^{\rm X}(z,z')$  with X=TE or X=TM, we use the representation
\be {{g}}_\Gamma^{\rm X}(z,z')=\frac{u_{\rm X}(z_<)v_{\rm X}(z_>)}{w}
\label{4.15}\ee
with
\bea v_{\rm X}(z)  &=&
\left(e^{ikz}+r_{\rm X}e^{-ikz}\right)\Theta(-z)+t_{\rm X}e^{iqz}\Theta(z),
\nn\\
u_{\rm X}(z) &=&
\bar{t}_{\rm X}e^{-ikz}\Theta(-z)+\left(e^{-iqz}+\bar{r}_{\rm X}e^{iqz}\right)\Theta(z),
\nn\\
w &=& -2iq{t}_{\rm X},
\label{4.16}\eea
and the well known reflection  coefficients,
\be r_{\rm TE}=\frac{q-k}{ q+k}, \ \
 r_{\rm TM}=\frac{\ep(\om) q-k}{\ep(\om) q+k}.
\label{4.17}\ee
The momenta are related by
\be \om^2=k_{||}^2+q^2, \ \ \ep(\om)\,\om^2=k_{||}^2+k^2.
\label{4.18}\ee
Now, since the polarization vectors are the same for the free space part $ \G^{(0)}_\om $ and for the addendum in \Ref{4.7}, we get the free space part from the above formulas with $r_{\rm X}\to 0$ and $t_{\rm X}\to 1$. Therefore, we get at $z=z'=a$,
\bea \Delta {{g}}_\Gamma^{\rm X}(a,a)&=&
\frac{\bar{t}_{\rm X}e^{iqa}\left(e^{-iqa}+{r}_{\rm X}e^{iqa}\right)}{-2iq \bar{t}_{\rm X}}
-\frac{1}{-2iq}
\nn\\&=&
\frac{{r}_{\rm X}}{-2iq}\, e^{2iqa}.
\label{4.19}\eea
Further, using \Ref{4.11}, we come to
\be \Delta \G^{(1)}_\om(\a,\a) = \int\frac{d\k_{||}}{(2\pi)^2}
    \sum_{\rm X=TE,TM} {\rm tr} \
    E_{\rm X}\frac{{r}_{\rm X}}{-2iq}\, e^{2iqa} E_{\rm X}^\dagger,
\label{4.20}\ee
where the trace is over the spatial structure.
Inserting into \Ref{4.6}, using \Ref{4.14} for the polarization vectors, we come under the trace to
\be  {\rm tr}\ln \bm{\hat{L}}(\om)
=  \ln (1-\al(\om) \om^2 \Delta G^{(1)})
\equiv \ln L(\om)\label{4.21}\ee
with
\be \Delta G^{(1)} = \int\frac{d\k_{||}}{(2\pi)^2}
\left( {r}_{\rm TE} +\frac{-q^2+k_{||}^2}{\om^2}  {r}_{\rm TM} \right)
\frac{e^{2iqa}}{-2i q}.
\label{4.22}\ee
We take \Ref{4.21} as definition for ${{L}}(\om)$, which we will use in the following. For convenience, we remind here the formulas \Ref{2.46} and \Ref{4.0},
\begin{widetext}
\be F= \int_0^\infty \frac{d\om}{\pi}\, \left(\frac{\om}{2}+T\ln\left(1-e^{-\beta\om}\right) \right) \, \pa_\om \delta(\om),
\ \
\delta(\om)=-\frac{1}{2i} \ln \frac{{{L}}(\om)}{{{L}}(-\om)},
\label{4.23}\ee
\end{widetext}
which, together with \Ref{4.21}, give the distance dependent part of the free energy.

In the following subsections we analyze the limiting cases and the transition to Matsubara representation. It will turn out, that this analysis goes much in parallel to that in the A-A configuration in Sects. \ref{T3.1} and \ref{T3.2}.

\subsection{\label{T4.1}Limiting cases}
\paragraph{Vanishing dissipation parameter and vacuum energy}
Here we restrict ourself to the second way mentioned in Sec. \ref{T3.1} using the $'\!\!+i0'$-prescription and take vanishing both, $\ga\to0$ and $\ga_0\to0$.
We start from representation \Ref{4.23} for $T=0$,
\be E_0= \left({F}_{|_{\ga=0}}\right)_{|_{T=0}} = \frac12 \int_0^\infty\frac{d\om}{\pi}\,\om
\pa_\om \frac{-1}{2i}\ln \frac{L(\om-i0))}{L(\om+i0)},
\label{4.24}\ee
where $L(\om)$ is given by \Ref{4.21} with $\ga=\ga_0=0$. Doing the Wick rotation and integrating by parts we get
\be E_0=\frac{1}{2\pi}\int_0^\infty d\xi\,\ln L(i\xi),
\label{4.25}\ee
where we assumed that criticality is absent. The explicit formulas are
\be L(i\xi)=1+\al(i\xi)\xi^2 \Delta G^{(1)}(i\xi)
\label{4.26}\ee
with %
\be \Delta G^{(1)}(i\xi) = \int\frac{d \k_{||}}{(2\pi)^2}
\left[r_{\rm TE}+\left(1-\frac{2\eta^2}{\xi^2}\right)r_{\rm TM}\right]
\frac{e^{-2a\eta}}{2\eta}.
\label{4.27}\ee
Under the Wick rotation, the momenta turn  into
$q=i\eta$ and $k=i\kappa$ and from relations \Ref{4.18} one comes to
\be \eta=\sqrt{\xi^2+k_{||}^2}, \ \
\kappa=\sqrt{(\ep(i\xi)-1)\xi^2+\eta^2}.
\label{4.28}\ee
Expressed in these momenta, the reflection coefficients are now
\be r_{\rm TE}=\frac{\eta-\kappa}{\eta+\kappa}, \ \
r_{\rm TM}=\frac{\ep(i\xi)\eta-\kappa}{\ep(i\xi)\eta+\kappa}.
\label{4.29}\ee
In the integration in \Ref{4.27}, it is meaningful to change the variable for $\eta$ and the final formula for $L(i\xi)$ is
\bea && L(i\xi) =  \label{4.30}\\&&
1+\frac{\al(i\xi)}{4\pi}\xi^2\int_\xi^\infty d\eta\
\left[r_{\rm TE}+\left(1-\frac{2\eta^2}{\xi^2}\right)r_{\rm TM}\right]
e^{-2a\eta}
\nn\eea
with $\al(i\xi)=\frac{\omp^2}{\xi^2+\Om_0^2}$. With these notations, $E_0$, \Ref{4.25}, gives the Casimir-Polder vacuum interaction energy in the A-W configuration. We mention, that this $E_0$ is non-perturbative in the polarizability.

We check the above formula by considering the case of an ideally conducting wall. In that case the reflection coefficients are $r_{\rm TE}=-1$, $r_{\rm TM}=1$ and the integration over $\eta$ can be carried out. Then the vacuum energy simplifies,
\bea E_0&=&\frac{1}{2\pi}\int_0^\infty d\xi\,\ln
\bigg(
1
 \label{4.31}\\ && \left.-\frac{2\al(i\xi)}{4\pi(2a)^3} (2+2(2a\xi)+(2a\xi)^2)
e^{-2a\xi}\right).
\nn\eea
Making here an expansion for small $\al(i\xi)$ or for large separation, $a$, in leading order one comes to
\be E_0=-\frac{3}{8\pi a^4} \frac{\al(0)}{4\pi},
\label{4.32}\ee
in agreement with Eq. (25) in \cite{casi48-73-360} or Eq. (16.28) in \cite{BKMM}.

At this place it should be mentioned that the onset of criticality follows from \Ref{4.30} to be at $\frac{\al(0)}{4\pi (2a)^3}>\frac14$ for $\Om_0\ne0$ and that criticality is always present for $\Om_0=0$. These are the same relations as in the A-A configuration.

\paragraph{Low temperature expansion}
We start from representation \Ref{3.19a} for the temperature dependent part of the free energy and Eq. \Ref{4.21} for $L(\om)$. However, we distinguish now between $\al(\om)$, \Ref{2.24}, for the atom (this is $i=0$ in \Ref{4.3}) and $\ep(\om)$, \Ref{4.3a}, for the half space, by giving an index $'0'$ to the parameters entering $\al(\om)$,
\be \al(\om)=\frac{4\pi e^2}{m({\color{red}-}\om^2-i\ga_0\om+\Om_0^2)}.
\label{4.32a}\ee
This way we will be able to trace the origin of different behavior.

Again, we need the expansion for $\om\to 0$. From \Ref{4.21} and \Ref{4.22} we get for $\Om_0\ne0$,
\be L(\om) = 1-\frac{\al(0)\omp^2}{\pi (2a)^3 (\omp^2+2\Om^2)}+O(\om),
\label{4.33}\ee
and for the phase,
\be \delta(\om)=
\left(\frac{2\ga}{\omp^2+2\Om^2}+\frac{\ga_0}{\Om_0^2}\right)  \om+O(\om^3).
\label{4.34}\ee
In the case $\Om_0=0$ we get
\bea L(\om) &=& \frac{ie^2\omp^2}{  2m a^3 \ga_0 (\omp^2+2\Om^2)}\frac{1}{\om}+O(1),
\nn\\
 \delta(\om) &=&
\left(\frac{2\ga}{\omp^2+2\Om^2}+\frac{1}{\ga_0}\right)  \om+O(\om^3).
\label{4.35}\eea
If, in addition, we have $\ga_0=0$, we get
\bea L(\om) &=& \frac{ie^2\omp^2}{  2m a^3   (\omp^2+2\Om^2)}\frac{1}{\om^2}+O(1),
\nn\\
 \delta(\om) &=&
 \frac{2\ga}{\omp^2+2\Om^2}  \, \om+O(\om^3).
\label{4.35a}\eea
In the derivation of these formulas, in \Ref{4.22} the expansion goes straight forward and, after that, the integration over $k_{||}$ is simple.

From \Ref{4.33} we see that a vanishing $\Om$ does not change the behavior essentially, whereas $\Om_0=0$ changes the behavior of $L(\om)$. The phases are not affected and start with first power in $\om$. Also we see that vanishing dissipation may increase only the power of the leading order.  Thus, using Eqs. \Ref{3.25} - \Ref{3.29}, we see a $ T^2$-behavior of the temperature dependent part of the free energy (at least) and, consequently, no violation of the third law is possible.
This holds also for a dissipation parameter vanishing with temperature due to the factor $\om^2$ in front of $\Delta G^{(1)}(\om)$ in \Ref{4.21}, see also the discussion in Sec. 16.3.3 in \cite{BKMM}. Problems, reported for dc conductivity (see Sects. 12.6.3 and 16.4.3 in \cite{BKMM}), where the TM polarization becomes important, we do not consider in this paper.
\subsection{\label{T4.2}Transition to Matsubara representation}
Starting point for the transition to Matsubara representation is Eq. \Ref{4.23} for the free energy. Further we can use a number of formulas from Sec. \ref{T3.2} dropping the index '$i$', namely \Ref{3.31} till \Ref{3.35}. For $\Om_0\ne0$ and without criticality, we get simply
\be F=T\sump \ln L(i\xi_l).
\label{4.36}\ee
For $\Om_0=0$, $L(0)$ is not finite and we need to go through the limiting procedure like in Sec. \ref{T3.2}. However, all calculations go completely in parallel. The difference starts from  \Ref{3.36}, in place of which  we get
\bea &&\frac{T}{2} \ln\left|L_1(i\ep)\right|
 \label{4.37}\\&&   =-\frac{T}{2}\ln\ep
    +\frac{T}{2}\ln\left(\frac{e^2\omp^2}{2ma^3 \ga_0 (\omp^2+2\Om^2)}\right)+O(\ep).
\nn\eea
For the contributions from the half circles first we need
\be \pa_\om \ln L(\om)=-\frac{1}{\om}+O(1)
\label{4.38}\ee
in place of \Ref{4.35} and similar to \Ref{3.39} we have now
\be F^{\rm h.circ.} =   \frac{T}{2}\ln(\beta\ep)+O(\ep).
\label{4.39}\ee
Again, the divergent logarithmic terms cancel as expected and we get finally
\bea F_i &=&=
\frac{T}{2}\ln\left(\frac{e^2\omp^2\beta}{2ma^3 \ga_0 (\omp^2+2\Om^2)}\right)
 \nn\\&&   +T\sum_{l=1}^\infty \ln\left|L_1(i\xi_l)\right|
    -T \ln\left|2\sin\frac{\beta\xi_*}{2}\right|.
\label{4.40}\eea
This way, for $\Om_0=0$ we have the formal substitution
\be  L(i\xi_{|_{l=0}}) \to
 \frac{e^2\omp^2\beta}{2ma^3 \ga_0 (\omp^2+2\Om^2)} .
\label{4.41}\ee
A similar result can be obtained if we have $\ga_0=0$ in addition.

\section{\label{T5}Wall-Wall configuration}
In this section we consider the W-W configuration. As set up in Sec. \ref{T4.2}, it consists of two half spaces filled with medium and a gap of width $a$ between them. This is just the setup of the Casimir effect. The free energy is given by Eq. \Ref{2.46} with $\delta(\om)$ given by Eq. \Ref{4.0} and $\L(\om)$ by \Ref{2.23}. We act in parallel to the preceding section and introduce a potential
\be V(\x)=V_1(z)+V_2(z)
\label{5.1}\ee
with
\bea V_1(z)&=&(\ep(\om)-1)\om^2\Theta(-z), \nn \\ V_2(z)&=&(\ep(\om)-1)\om^2\Theta(z-a), \ \
\label{5.2}\eea
where $\ep(\om)$ is given by \Ref{4.3a}. Next, we consider
\be {\rm tr}\ln \bm{\hat{L}}(\om)
={\rm tr}\ln\left(1-(V_1(z)+V_2(z))\hG^{(0)}_\om\right).
\label{5.3}\ee
This expression can be split into
\bea {\rm tr}\ln\bm{\hat{L}} &=& {\rm tr}\ln\left(1-V_1(z)\hG^{(0)}_\om\right)
                +{\rm tr}\ln\left(1-V_2(z)\hG^{(0)}_\om\right)
      \nn\\   &&       +{\rm tr}\ln(1-{\cal M}),
\label{5.4}\eea
where
\bea {\cal M}& =&
                \left(1-V_1\hG^{(0)}_\om\right)^{-1}\bm{\hat{V}}_1\hG^{(0)}_\om
                \left(1-V_2\hG^{(0)}_\om\right)^{-1}\bm{\hat{V}}_1\hG^{(0)}_\om ,
\nn \\ &=&          \bm{\hat{T}}_1\hG^{(0)}_\om\bm{\hat{T}}_2\hG^{(0)}_\om.
\label{5.5}\eea
Here, $\bm{\hat{T}}_1$ and $\bm{\hat{T}}_2$ are the T-operators for the potentials $V_1(\x)$ and $V_2(\x)$ taken separately. Further, in deriving this equation, formulas like \Ref{2.47} and \Ref{2.48} were used as well as the cyclic property of the trace. Eq. \Ref{5.4} is nothing else then the well known transition to the 'TGTG'-formula \cite{kenn06-97-160401}. Equivalent formulas can be found also in the so-called scattering approach \cite{rahi09-80-085021}. The first two terms in \Ref{5.4} give contributions to the free energy, which will not depend on the width of the gap and we drop them. This way, we have to consider
\be {\rm tr}\ln\bm{\hat{L}} =  {\rm tr}\ln(1-{\cal M}),
\label{5.6}\ee
which is a well known expression in connection with the Casimir effect, see for instance Sec. 10.1.2 in \cite{BKMM}. At this place it should be mentioned, that the known formulas are very similar to that derived here, which, however, follow from the heat bath approach.

Continuing from Eq. \Ref{5.6}, we can use the known explicit formulas, see for example Sec. 12.1 in \cite{BKMM} (however, with slightly different notations). This way we get
\be {\rm tr}\ln\bm{\hat{L}}=\int\frac{d\k_{||}}{(2\pi)^2}\,
    \sum_{\rm X=TE,TM}\ln\left(1-r_{\rm X}^2 \,e^{2iaq}\right),
\label{5.7}\ee
where the reflection coefficients $r_{\rm TE}$ and $r_{\rm TM}$ are defined in \Ref{4.17} and the momenta $q$ and $k$ in \Ref{4.18} with $\ep(\om)$ defined in \Ref{4.3a}.

For the convenience of the following, we rewrite the above representation of the free energy in the form
\be F=\frac{1}{4\pi^2}\int_0^\infty d\om\,
    \left(\frac{\om}{2}+T\ln\left(1-e^{-\beta\om}\right)\right)
    \pa_\om \phi(\om)
\label{5.8}\ee
with
\be \phi(\om)=\frac{-1}{2i}\left(\varphi(\om)-\varphi(\om)^*\right),
\label{5.9}\ee
and
\be \varphi(\om) = \int_0^\infty dk_{||} k_{||}
    \sum_{\rm X=TE,TM}\ln\left(1-r_{\rm X}^2 \,e^{2iaq}\right).
\label{5.10}\ee
With formulas \Ref{5.8}-\Ref{5.10} we have, in fact, a representation of the Lifshitz formula with finite dissipation $\ga$ in terms of real frequencies. In fact, it is not really new. A similar formula was obtained in \cite{bord14-981586}, however starting from the Matsubara representation with Drude permittivity inserted, and applying the Abel-Plana formula.  (Eq. (58) in \cite{bord14-981586}). Eqs. (62) and (63) in \cite{bord14-981586} correspond to the above \Ref{5.8}-\Ref{5.10}, however with slightly different notations. Also we mention, that the analysis done in \cite{bord14-981586} contains nearly all calculations which are needed here to investigate the limiting cases of the free energy \Ref{5.8} and its relation to the Matsubara representation.

We mention that representation \Ref{5.8} for the free energy can also be rewritten using the same notations as in \Ref{2.46} and \Ref{2.45}, with a different definition for $\bm{\hat{T}}(\om)$ and the trace. However, because we take over a number of formulas from \cite{bord14-981586}, it is meaningful also to use the notations from there.
\subsection{\label{T5.1}Limiting case $T\to0$}
In fact, the limiting case for $T\to0$ was calculated in \cite{bord14-981586} (and earlier, see literature cited therein). For this reason we restrict ourself here to a short review of the methods used and display the results.

The first step of the method is the same as in Sec. \ref{T3.1}, Eqs. \Ref{3.25a}-\Ref{3.28}. One derives an expansion of $\delta(\om)$ for $\om\to0$ and inserting this expansion into the free energy gives its expansion for $T\to0$. In order to get the expansion for $\om\to0$, it is meaningful to divide the integration region in \Ref{5.10} into two regions, (a) with $\om>k_{||}$ and (b) with $\om<k_{||}$, splitting $\varphi(\om)$ accordingly,
\be \varphi(\om)=\varphi_a(\om)+\varphi_b(\om),
\label{5.11}\ee
with
\bea \varphi_a(\om) &=& \int_0^\om dk_{||} k_{||}
    \sum_{\rm X=TE,TM}\ln\left(1-r_{\rm X}^2 \,e^{2iaq}\right),\nn \\
    \varphi_b(\om) &=& \int_\om^\infty dk_{||} k_{||}
    \sum_{\rm X=TE,TM}\ln\left(1-r_{\rm X}^2 \,e^{2iaq}\right).\nn\\
\label{5.12}\eea
In this division, the first part, $\varphi_a(\om)$, is irrelevant for the leading orders in $\om$ since it decreases not slower then $\sim\om^2$ because of the integration interval. In $\varphi_b(\om)$, where we have $q=\sqrt{-\om^2+k_{||}^2}\equiv i \eta$ with $\eta=\sqrt{\om^2+k_{||}^2}$ real, it is meaningful to change the integration variable for $\eta$. Then the relevant formulas are
\bea k=i\kappa,\ \ \kappa=\sqrt{\eta^2-(\ep(\om)-1)\om^2},\nn\\
    r_{\rm TE}=\frac{\eta-\kappa}{\eta+\kappa}, \ \
    r_{\rm TM}=\frac{\ep(\om)\eta-\kappa}{\ep(\om)\eta+\kappa}
\label{5.13}\eea
and we have
\be \varphi_b(\om) = \int_0^\infty d\eta \,\eta\,
    \sum_{\rm X=TE,TM}\ln\left(1-r_{\rm X}^2 \,e^{-2a\eta}\right).
\label{5.14}\ee
We insert the permittivity \Ref{4.3a} and $N(\om)$, \Ref{2.24}, into $\kappa$, \Ref{5.13},
\be \kappa=\sqrt{\eta^2-\omp^2\frac{\om^2}{-\om^2-i\ga\om+\Om^2}}.
\label{5.15}\ee
Now, for $\Om\ne0$, one may obtain a power series expansion in $\om$ for $\om\to0$ and this case is not really interesting. It is largely equivalent to fixed permittivity, considered in Sec. 4.2 in \cite{bord14-981586}. Therefor we continue with the case $\Om=0$. In this case, $\ep(\om)$, \Ref{4.3a},  is that of the Drude model,
\be \ep^{\rm Dr}(\om)=1-\frac{\omp^2}{\om(\om+i\ga)}.
\label{5.16}\ee
The calculation of $\varphi_b(\om)$ for $\om\to0$ requires the expansion of integrals like that in \Ref{5.12}. Partly, corresponding formulas can be found in literature. A most complete expansion is given in the App. A in \cite{bord14-981586}. Using these results, the
following expansions were obtained. From
Eqs. (165) and (171) in \cite{bord14-981586} we get,
\bea \phi_{\rm TE}(\om) &=& \frac{2\ln 2-1}{2}\frac{\omp^2}{\ga^2}\,\om - \frac{1}{6\sqrt{2}}\left(\frac{\omp^2}{\ga}\right)^3\om^{3/2}+O(\om^2),\nn\\
\phi_{\rm TM}(\om) &=& -\frac{4\pi^2\ga}{3\omp^2}\om +O(\om^2),
\label{5.17}\eea
and from (275) in \cite{bord14-981586}
\bea  \Delta_T F &=&
\left(\frac{(2\ln 2-1)\omp^2}{\ga}-\frac{2\pi^2\ga}{3a^2\omp^2}\right)\frac{T^2}{48}
\nn\\&&-\frac{\zeta(5/2)\omp^2}{16\sqrt{2}\ga^{3/2}}a T^{5/2}+O(T^3)
\label{5.18}\eea
follows.
This way, the expansion of the free energy starts with $T^2$ and the third law of thermodynamics is respected.

\subsection{\label{T5.2}Limiting case $\ga\to0$ and relation to the plasma model}
For $\Om=0$, the limit $\ga\to0$ of vanishing dissipation parameter is the most interesting one since it causes the problems with the third law. There are two aspects. The one is the limit $\ga\to0$ at fixed $T$ and the other is the limit $T\to0$ with a temperature dependent dissipation parameter $\ga(T)$, having the property
\be \ga(T)\raisebox{-4pt}{${\sim\atop T\to0}$}\ga_1 T^\al, \ \ (\al>1),
\label{5.19}\ee
where $\ga_1$ is some constant, i.e., vanishing faster than the first power of the temperature.

On a formal level,  putting $\ga = 0$ turns the permittivity \Ref{5.16} of the Drude model into that of the plasma model,
\be \ep^{\rm pl}(\om)=1-\frac{\omp^2}{\om^2},
\label{5.20}\ee
and the same holds for the reflection coefficients.
However, the free energy does not do the same and an additional contribution appears.
This can be seen in the following way. First, we remark that the additional contribution appears from the TE polarization  to the function $\varphi_b(\om)$ in \Ref{5.14},
\be \varphi_b^{\rm TE}(\om) = \int_0^\infty d\eta \,\eta\,
     \ln\left(1-r_{\rm TE}^2 \,e^{-2a\eta}\right).
\label{5.21}\ee
Here, the dissipation parameter enters through the momentum $\kappa$, \Ref{5.15}, which using \Ref{5.16} now reads
\be \kappa=\sqrt{\eta^2+\frac{\om_p^2 \om}{\om+i\ga}}.
\label{5.22}\ee
This way, the dependence on $\ga$ enters only through the quotient $\om/\ga$ and we define
\be \varphi_b^{\rm TE}(\om) \equiv \psi\left(\frac{\om}{\ga}\right).
\label{5.23}\ee
%
%
%
%
For considering $\ga\to 0$, we make the substitution $\om\to\om \ga$ in the corresponding part of the free energy \Ref{5.8}, and get
\bea \Delta_T {F_b^{\rm TE}}&=&\frac{T}{4\pi^2}\int_0^\infty d\om\,
    \ln\left(1-e^{-\beta\ga\om}\right)
 \nn\\&&  \cdot \pa_\om \frac{1}{2i}\left(\psi(\om)-\psi(\om)^*\right).
\label{5.24}\eea
Now, because $\psi(\om)$ is decreasing for $\om\to0$, as follows from \Ref{5.17},  and for $\om\to\infty$, as can be shown easily, we can take $\ga\to0$ in the logarithm,
$\ln\left(1-e^{-\beta\ga\om}\right) =\ln \om +\ln(\beta\ga)+\dots\ $. The term $\ln(\beta\ga)$ does not contribute to $\Delta_T F$ and from the first term we get
\be \Delta_T F \raisebox{-4pt}{$=\atop\ga\to0$}\
\frac{T}{4\pi^2}\int_0^\infty d\om\,
    \ln\om\,
    \pa_\om \frac{1}{2i}\left(\psi(\om)-\psi(\om)^*\right).
\label{5.25}\ee
The integral over $\om$ can be carried out, as shown in App. B in \cite{bord14-981586}, and one arrives at (see Eq. (183) in \cite{bord14-981586})
\be \lim_{\ga\to0}\Delta_T F=\Delta_T F^{\rm pl}+{\cal F}_1 T
\label{5.26}\ee
%
with
\be {\cal F}_1 = \frac{-1}{4\pi^2}\int_0^\infty d\eta\,\eta\,
    \ln\left(1- \left( {r_{\rm TE}^{\rm pl}} \right) ^2\,e^{-2a\eta}\right),
\label{5.27}\ee
where
\be r_{\rm TE}^{\rm pl}=\frac{\eta-\sqrt{\eta^2+\omp^2}}{\eta+\sqrt{\eta^2+\omp^2}}
\label{5.28}\ee
is the reflection coefficient of the TE polarization in the plasma model and $\Delta_T F^{\rm pl}$ is the free energy of the plasma model, i.e., the free energy with $\ga=0$ taken from the very beginning.

The observation, that the limit $\ga\to0$, delivers an additional contribution linear in T, was made already in \cite{bord11-71-1788} (see also the discussion in Sec. 4.4.2 in \cite{bord14-981586}, or in Sec. 14.1 in \cite{BKMM}), for the representation of the free energy in terms of Matsubara frequencies, which we will consider in the next subsection. Here we mention only that now this behavior follows also within the heat bath approach.

Using the above formulas, it is  easy to consider the second aspect, i.e., the limit $T\to0$ with $\ga(T)$ decreasing as shown in \Ref{5.19}. The relevant contribution comes again from \Ref{5.24}. Inserting \Ref{5.19} we get
\bea \Delta_T {F_b^{\rm TE}}&=&\frac{T}{4\pi^2}\int_0^\infty d\om\,
    \ln\left(1-\exp\left({- \ga_1\om T^{\al-1}}\right)\right)
 \nn\\&&\cdot   \pa_\om \frac{1}{2i}\left(\psi(\om)-\psi(\om)^*\right).
\label{5.29}\eea
Now, for $\al>1$, $T\to0$ calls like $\ga\to0$ in \Ref{5.23} for an expansion of the logarithm and using the same arguments one comes to
\be \Delta_T F = T {\cal F}_1+O(T^2),
\label{5.30}\ee
where ${\cal F}_1$ is the same as in \Ref{5.24}. This way, a temperature dependent dissipation parameter with the property \Ref{5.19} delivers for small $T$ a linear contribution, which violates the third law.

Let us consider the relation to the free energy calculated within the plasma model in more detail. We have seen  that an additional contribution comes from the TE polarization and the frequency region (b), defined in \Ref{5.12}, i.e., from a region where $\om\le k_{||}$ holds. In general, as discussed in detail in \cite{bord12-85-025005}, see especially Fig.2, there are frequency regions corresponding to wave guide and scattering modes. These have $\om>k_{||}$ (region (a)). For $\om<k_{||}$ (region (b)), there are surface modes, however only in the TM polarization. This way, the additional contribution comes from the frequency region (b), where in the plasma model there are no modes.

We add a comment on the limit $\ga\to0$. Before the substitution $\om\to\ga\om$, resulting in \Ref{5.24}, the contribution from the region (b) to the temperature dependent part of the free energy, following from \Ref{5.8}, reads
\bea \Delta_T {F_b^{\rm TE}}&=&\frac{T}{4\pi^2}\int_0^\infty d\om\,
    \ln\left(1-e^{-\beta\om}\right)
 \nn\\&&\cdot   \pa_\om \frac{1}{2i}\left(\psi\left(\frac{\om}{\ga}\right)-\psi\left(\frac{\om}{\ga}\right)^*\right).
\label{5.30a}\eea
If we put $\ga=0$ here directly in the integrand we get zero since $\psi(\infty)$ is real. This can be seen from \Ref{5.22} which turns into $\kappa=\sqrt{\eta^2+\omp^2}$, which is the same as in \Ref{5.28}. However, as we have seen above, this result is incorrect. Indeed, interchanging the limit $\ga\to0$ and the integration in the representation \Ref{5.30a} is not allowed in distinction to the representation \Ref{5.24}.

\subsection{\label{T5.3}Transition to Matsubara representation}
In the W-W configuration, the transition to Matsubara representation can be done the same way is in the preceding two section. But now, it is even easier since $\phi_{\rm X}(\om=0)=0$ (see \Ref{5.17}) and no problems may appear in dividing the integral in \Ref{5.8} into two integrals according to \Ref{5.9}. This way, from \Ref{5.8}-\Ref{5.10}, we come to the known formulas for the free energy in Matsubara representation, i.e., to the Lifshitz formula with finite dissipation $\ga$. These formulas read
\be F=\frac{1}{4\pi^2}T\sump \varphi(i\xi_l), \ \ (\xi_l=2\pi T L),
\label{5.31}\ee
with
\be \varphi(i\xi)=\int_{\xi_l}^\infty d\eta \, \eta \, \sum_{\rm X=TE,TM}
    \ln\left(1-r_{\rm X}^2 e^{-2a\eta}\right)
\label{5.32}\ee
and the reflection coefficients
\be r_{\rm TE}=\frac{\eta-\kappa}{\eta+\kappa}, \
r_{\rm TM}=\frac{\ep(i\xi)\eta-\kappa}{\ep(i\xi)\eta+\kappa},
\label{5.33}\ee
which are the same as in \Ref{5.13} except for  momentum and permittivity,
\be \kappa=\sqrt{\eta^2+\omp^2\frac{\xi^2}{\xi^2+\xi \ga+\Om^2}}, \ \ \ep(i\xi)=1+\frac{\omp^2}{\xi^2+\xi \ga+\Om^2}.
\label{5.34}\ee
Here several comments are in order. In the Lifshitz formula \Ref{5.31}, criticality never appears. This can be seen in \Ref{5.32}, where we have $r_{\rm X}\le1$ since all quantities entering \Ref{5.33} are nonnegative.
For $\Om=0$, which makes the permittivity \Ref{5.34} to that of a metal,  for $l=0$ a peculiarity appears. In general, for $\xi\to0$ we note $\ep(i\xi)\to\infty$, and the reflection coefficients turn into
\be r_{\rm TE}\to 0,\ \ r_{\rm TM}\to 1,
\label{5.35}\ee
i.e., only the reflection coefficients for the TM polarization turns into that of an ideal conductor, and that of the TE polarization turns into zero. This way, the contribution from the TE polarization to the zeroth Matsubara frequency is missing. This is counterintuitive, especially, since $l=0$ determines the behavior of the free energy for $T\to\infty$. This problem is well known, and discussed, for instance, in Chapt. 14.1 in \cite{BKMM}. A similar problem was observed earlier for fixed permittivity $\ep$ (for $\ep\to\infty$ we have also \Ref{5.35}), and 'cured' by the Schwinger-prescription in \cite{schw78-115-1}.
Here we have to state, that this problem appears in the heat bath approach too.

We mention, that the linear term ${\cal F}_1T$ in \Ref{5.26} can be seen in Matsubara representation easier since it simply describes the difference between Drude model for $\ga\to0$ and plasma model at $\ga=0$,
which in Matsubara representation shows up in the zeroth Matsubara frequency, $l=0$, and $\xi_0=0$.
%
As seen from \Ref{5.28}, the reflection coefficient for the plasma model does not depend on $\xi$ and $\xi=0$ gives $r_{\rm TE}\ne0$, whereas from \Ref{5.33} and \Ref{5.34}, for $\Om=0$ and $\ga\ne0$, \Ref{5.35} follows for $\xi=0$. This way, $l=0$
turns \Ref{5.29} just into ${\cal F}_1T$ with ${\cal F}_1$ given by \Ref{5.27}.

\section{\label{T6}Conclusions}
In the foregoing sections we considered dipoles interacting with the electromagnetic field and with heat baths. The aim is to describe, in this system, dissipation from first principles. Indeed, and this was in principle known before, this is possible using known formulas. We got an general representation, Eq. \Ref{2.46}, for the free energy.

This representation is similar the  'remarkable formula' in \cite{ford85-55-2273} and one may expect  that it reflects some more general underlying structure. For our system, we have with \Ref{2.45} a specific expression in terms of the T-operator of the effective equation \Ref{2.27} for the electric field.

We applied the general formula to three configurations, A-A and A-W which correspond to the Casimir-Polder force and to W-W, which corresponds to the Casimir effect. In each case, we obtain a representation of the free energy in terms of an integral over real frequencies in the presence of dissipation. For the A-A and A-W configurations, we obtain formulas which are  nonperturbative in the polarizability, i.e., include all orders. Of course, from here the known expansions \Ref{3.19} and \Ref{4.32}, for small polarizability or large separation, follow. It is to be mentioned, that in this sense the W-W configuration, i.e., the Lifshitz formula, is always nonperturbative.

In the nonperturbative treatment of the A-A and A-W configurations, criticality may appear. As a result, the transition to the Matsubara representation becomes modified.
Another modification of the Matsubara representation occurs if the intrinsic frequency of the atoms in the A-A and A-W configurations vanishes. This modification results in a change of the contribution from the zeroth Matsubara frequency, Eqs. \Ref{3.43},\Ref{3.44} and \Ref{4.41}.

The main motivation for the present paper was to look on the problems with thermodynamics known in connection with the Casimir effect for temperature dependent dissipation parameter, decreasing faster than the first power of the temperature. First of all, the dissipation parameter $\ga$, as it appears in Eq. \Ref{2.11} from the heat bath, does not depend on temperature (it may depend on frequency), so to say, 'by construction'. The temperature dependence comes in solely through the averages \Ref{2.12}. We show that in all cases, for a constant $\ga$, the laws of thermodynamics are respected. If however, inserting 'by hand' a temperature dependence into the dissipation parameter as given by eq.  \Ref{5.19}, the known problems appear; now also in the heat bath approach. We conclude that the heat bath approach, at least in the present form, is insufficient to solve the mentioned problems. In calling for another approach one should bear in mind that the temperature dependence \Ref{5.19} follows from the behavior of real metals and, this way, cannot be ignored.

\bibliographystyle{unsrt}
\bibliography{C:/Users/bordag/WORK/Literatur/bib/papers,C:/Users/bordag/WORK/Literatur/bib/libri,C:/Users/bordag/WORK/Literatur/Bordag}

\begin{thebibliography}{10}

\bibitem{mahanty76}
J.~Mahanty and B.~W. Ninham.
\newblock {\em Dispersion forces}.
\newblock Academic, New York, 1976.

\bibitem{casi48-51-793}
H.~B.~G. Casimir.
\newblock On the attraction between two perfectly conducting plates.
\newblock {\em Proc. K. Ned. Akad. Wet.}, 51:793--795, 1948.

\bibitem{lifs56-2-73}
E.~M. Lifshitz.
\newblock The theory of molecular attractive forces between solids.
\newblock {\em Zh.Eksp.Teor.Fiz.}, 29:94, 1956.
\newblock [Sov.Phys.JETP. {\bf 2}, 73 (1956)].

\bibitem{bara75-116-5}
Yu.~S. Barash and V.~L. Ginzburg.
\newblock {Electromagnetic Fluctuations in Matter and Molecular (van der Waals)
  Forces Between Them}.
\newblock {\em Usp.~Fiz.~Nauk}, 116:5, 1975.
\newblock [Sov.Phys. Usp. {\bf 18}, 306 (1975)].

\bibitem{BKMM}
M.~Bordag, G.~L. Klimchitskaya, U.~Mohideen, and V.~M. Mostepanenko.
\newblock {\em Advances in the Casimir Effect}.
\newblock Oxford University Press, Oxford, 2009.

\bibitem{klim09-81-1827}
G.~L. Klimchitskaya, U.~Mohideen, and V.~M. Mostepanenko.
\newblock {The Casimir force between real materials: Experiment and theory}.
\newblock {\em Rev.~Mod.~Phys.}, {81}:{1827}, {2009}.

\bibitem{beze02-65-052113}
V.~B. Bezerra, G.~L. Klimchitskaya, and V.~M. Mostepanenko.
\newblock {Thermodynamical Aspects of the Casimir Force Between Real Metals at
  Nonzero Temperature}.
\newblock {\em Phys.~Rev.~A}, {65}:{052113}, {2002}.

\bibitem{bimo16-93-184434}
G.~Bimonte, D.~L\'opez, and R.~S. Decca.
\newblock {Isoelectronic determination of the thermal Casimir force}.
\newblock {\em Phys.~Rev.~B}, 93:184434, 2016.

\bibitem{kupi92-46-2286}
D.~Kupiszewska.
\newblock {Casimir Effect in Absorbing Media}.
\newblock {\em Phys.~Rev.~A}, {46}:{2286}, {1992}.

\bibitem{rosa11-84-053813}
F.~S.~S. Rosa, D.~A.~R. Dalvit, and P.~W. Milonni.
\newblock {Electromagnetic energy, absorption, and Casimir forces. II.
  Inhomogeneous dielectric media}.
\newblock {\em Phys.~Rev.~A}, {84}:{053813}, {2011}.

\bibitem{lomb11-84-052517}
Fernando~C. Lombardo, Francisco~D. Mazzitelli, and Adrian~E. Rubio~Lopez.
\newblock {Casimir Force for Absorbing Media in an Open Quantum System
  Framework: Scalar Model}.
\newblock {\em Phys.~Rev.~A}, 84:052517, 2011.

\bibitem{lope17-28-025009}
Adri\'an~E. Rubio~L\'opez.
\newblock Quantum vacuum fluctuations in presence of dissipative bodies:
  Dynamical approach for nonequilibrium and squeezed states.
\newblock {\em Phys.~Rev.~D}, 95:025009, Jan 2017.

\bibitem{brau17-190-237}
M.~A. Braun.
\newblock {The Casimir Energy in a Dispersive and Absorptive Medium in the Fano
  Diagonalization Approach}.
\newblock {\em Theor. Mat. Phys.}, {190}({2}):{237--250}, {2017}.

\bibitem{casi48-73-360}
H.~B.~G. Casimir and D.~Polder.
\newblock {The Influence of Retardation on the London-van der Waals Forces}.
\newblock {\em Phys.~Rev.}, 73:360, 1948.

\bibitem{D17-1}
M.~Bordag.
\newblock {Vacuum and Thermal Energies for two Oscillators Interacting Through
  a Field}.
\newblock arXiv: 1707.06214, to appear in "Theoretical and Mathematical
  Physics".

\bibitem{ford85-55-2273}
G.~W. Ford, J.~T. Lewis, and R.~F. O'Connell.
\newblock {Quantum Oscillator in a Blackbody Radiation Field}.
\newblock {\em Phys.~Rev.~Lett.}, 55:2273--2276, 1985.

\bibitem{intr12-86-062517}
Francesco Intravaia and Ryan Behunin.
\newblock {Casimir Effect as a Sum over Modes in Dissipative Systems}.
\newblock {\em Phys.~Rev.~A}, {86}, {2012}.

\bibitem{pass12-85-062109}
Roberto Passante, Lucia Rizzuto, and Salvatore Spagnolo.
\newblock {Harmonic Oscillator Model for the Atom-Surface Casimir-Polder
  Interaction Energy }.
\newblock {\em Phys.~Rev.~A}, 85:062109, 2012.

\bibitem{berm14-89-022127}
P.~R. Berman, G.~W. Ford, and P.~W. Milonni.
\newblock {Nonperturbative Calculation of the London-van der Waals Interaction
  Potential}.
\newblock {\em Phys.~Rev.~A}, {89}, {2014}.

\bibitem{Akhiezer1985}
A.I. Akhiezer and I.A. Akhiezer.
\newblock {\em Electromagnetism and Electromagnetic Waves}.
\newblock 2Vysshaya Shkola, Moscow, 1985.
\newblock [in Russian].

\bibitem{milton98}
Julian Schwinger, Lester~L. {DeRaad, Jr.}, Kimbal~A. Milton, and Wu-yang Tsai.
\newblock {\em {Classical Electrodynamics}}.
\newblock {Perseus Books, Reading, Massachusetts}, 1998.

\bibitem{rosa10-81-033812}
F.~S.~S. Rosa, D.~A.~R. Dalvit, and P.~W. Milonni.
\newblock {Electromagnetic Energy, Absorption, and Casimir Forces: Uniform
  Dielectric Media in Thermal Equilibrium}.
\newblock {\em Phys.~Rev.~A}, {81}:{033812}, {2010}.

\bibitem{ford88-37-4419}
G.~W. Ford, J.~T. Lewis, and R.~F. O'Connell.
\newblock {Quantum Langevin Equation}.
\newblock {\em Phys.~Rev.~A}, 37:4419--4428, 1988.

\bibitem{bord15-91-065027}
M.~Bordag and J.M. Munoz-Castaneda.
\newblock {Dirac Lattices, Zero-Range Potentials and Self Adjoint Extension}.
\newblock {\em Phys.~Rev.~D}, 91:065027, 2015.

\bibitem{sche08-58}
Stefan Scheel and Stefan~Yoshi Buhmann.
\newblock {Macroscopic Quantum Electrodynamics --- Concepts and Applications}.
\newblock {\em {A}cta {P}hysica {S}lovaca}, 58:675, 2008.

\bibitem{kenn06-97-160401}
Oded Kenneth and Israel Klich.
\newblock {Opposites Attract - A Theorem About The Casimir Force}.
\newblock {\em Phys.~Rev.~Lett.}, 97:160401, 2006.

\bibitem{rahi09-80-085021}
Sahand~Jamal Rahi, Thorsten Emig, Noah Graham, Robert~L. Jaffe, and Mehran
  Kardar.
\newblock {Scattering Theory Approach to Electrodynamic Casimir Forces}.
\newblock {\em Phys.~Rev.~D}, 80:085021, 2009.

\bibitem{bord14-981586}
M.~Bordag.
\newblock {Low Temperature Expansion in the Lifshitz Formula}.
\newblock {\em Adv.~Math.~Phys.}, page 981586, 2014.

\bibitem{bord11-71-1788}
M.~Bordag.
\newblock {Drude Model and Lifshitz Formula}.
\newblock {\em Eur.~Phys.~J.~C}, 71:1788, 2011.

\bibitem{bord12-85-025005}
M.~Bordag.
\newblock {Electromagnetic Vacuum Energy for two Parallel Slabs in Terms of
  Surface, Wave Guide and Photonic Modes}.
\newblock {\em Phys.~Rev.~D}, {85}:{025005}, 2012.

\bibitem{schw78-115-1}
J.~Schwinger, L.L. DeRaad, Jr., and K.A. Milton.
\newblock {Casimir Effect in Dielectrics}.
\newblock {\em Ann. Phys.}, 115:1--23, 1978.

\end{thebibliography}
\end{document}